\documentstyle{article}
\title{Heterogeneous condensation  on   several types of
centers in dynamic conditions}
\author{V.Kurasov}
\date{Victor.Kurasov@pobox.spbu.ru}

\begin{document}
\maketitle
\begin{abstract}

A system of a metastable phase with  several sorts of the heterogeneous centers is
considered.
An analytical theory for the process of
condensation in such a system is constructed
in dynamic
conditions.
The free energy of formation of the critical embryo
 is assumed to be known in
the macroscopic  approach as well as the  energy of solvatation.
The recurrent procedure of the establishing of the
characteristic times and lengths is presented.
The semiempirical method of the description of  the
period  of  essential  formation  of  new  phase  embryos  on   the
heterogeneous
centers of some sort is given.

\end{abstract}

\pagebreak

\section{Introduction}

The theory considered here
will be  based  on  the  capillary  approximation  of
the height of the activation
barrier.
  This publication can be regarded as the direct continuation of the
publications \cite{aero1}, \cite{aero2} and is going to fulfil the program
announced in \cite{aero1}. All necessary bibliographical cites
can be found in \cite{aero1}.

It is necessary to stress that in the nature the more spread conditions
are the external conditions of the dynamic type.
In such conditions the external action on the volume
of the system, the temperature and the pressure are changed in some
continuous
smooth way.
The process of condensation violates these characteristics.
But this process itself takes place under the action of external
influence on the
 metastability of the system. These conditions were considered at first
by Raiser \cite{Rai}. The theory of
the heterogeneous condensation on the centers of one sort was done in
\cite{Novosib}.

Practically in
the condensing system there are
several types of the heterogeneous centers of
the different nature (see \cite{aero1}).

The theory
when the processes of the
homogeneous and heterogeneous nucleation are taken into account
in the dynamic conditions was constructed in \cite{Deponall}.
The problem of construction of the theory for condensation
on the heterogeneous centers of several sorts still remains actual. Here such
a theory will be constructed.

Here and in the further considerations all energy-like
values are expressed in the units of the thermal energy $k_{b}T$
($k_{b}$ is Bolzman constant, $T$ is absolute temperature).
The theory presented here will be valid also for
the pure homogeneous process.

We shall use
the following physical assumptions analogous to \cite{aero1}, \cite{aero2}:
\begin{itemize}
\item
the thermodynamic description of the critical embryo,
\item
the random homogeneous space distribution of the
heterogeneous centers,
\item
the free-molecular regime of the droplets growth,
\item
the homogeneous
external conditions for the   temperature and for the pressure,
\item
rather a  high
activation barrier of condensation,
\item
the absence of the thermal effects.
\end{itemize}
As far as the most interesting characteristics of this process are the
 numbers of the heterogeneously
formed droplets of the different types
we shall estimate the accuracy of the theory
by
 the error of the obtained solutions for
 these values.
The unit volume is considered.

We assume the total number of the heterogeneous centers to be constant in
time.

The indexes "*",  "**" below the values   may  refer to some characteristic
times $t_*$ and $t_{**}$, correspondingly.

The theory of the process appear to be rather long and complex.
So we shall use some already known constructions
\cite{Novosib}, \cite{Deponall}
as some blocks in this theory.
Recall that the heterogeneous condensation under the action of the linear source
of the metastability was considered in \cite{Novosib}. The same problem for
the source of the "square" form was studied in \cite{Deponall}.

Suppose that in the system there are several sorts of the heterogeneous centers
distributed homogeneously in the space volume. We
shall mark the total number of the heterogeneous centers by
$\eta_{tot\ i}$ where $i$ corresponds to some sort of heterogeneous centers.
The real values of
the free heterogeneous centers which
may be solvatated but aren't occupied by
the super-critical embryos are marked by $\eta_{i}$. The index $i$ or
$j$ below the
value
marks the sort of the heterogeneous center. The absence of this index points that
the formula is valid for an arbitrary sort of the heterogeneous centers.
          The density of the molecules in the
equilibrium vapor  is marked
by \( n_{\infty} \), the density of the molecules in the
real vapor in the system  is marked by \( n\).
 The power of the metastability will be characterized by the
value of the supersaturation
$$ \zeta = \frac{ n - n_{\infty} }{ n_{\infty} } $$
We shall define the super-critical embryos as the "droplets".
Every droplet is described by the number of the molecules inside the droplet
\( \nu \) , or by the linear size \( \rho = \nu^{1/3} \) .
Due to the free-molecular regime of
the vapor consumption  we have
$$ \frac{d\rho}{dt} = \zeta \alpha \tau^{-1} $$
where \( \alpha \) is the condensation coefficient and \( \tau \) is
some characteristic
 time between collisions of the given molecule of the
vapor with the molecules obtained from the gas kinetic theory.

Let us introduce some size \( z \)
according to
\begin{equation}
\label{2}
z = \int_{t_*}^{t} \zeta \alpha \tau^{-1} dt'
\end{equation}
where $t_*$ is some characteristic time. The choice of $t_*$ is more
complex
than in \cite{Novosib}. For every sort of the heterogeneous centers there
will be some special $t_{*\ i}$. It is chosen as the moment until which
the half of the droplets formed on the centers of the given sort is already
appeared. In such a case the variable $z$ will be
marked as $z_i$.

Until the coalescence \cite{15}, \cite{16}
which isn't considered here
equation (\ref{2}) ensures the growth of \( z \) in time and can be
inverted
\begin{equation} \label{3}
t(z) =  \alpha^{-1} \int_{0}^{z} \frac{\tau dz}{\zeta(z)} + t_*
\end{equation}
Hence, all values dependent on time become the values dependent on
\( z\) and the relative size \( x=z-\rho \) can be introduced.
For $z_i$  the values $x_i$ are introduced by the same procedure.

During the whole evolution the droplet has one and the same
value of the variable \( x_i \).
Considering \( t(x) \) (or $t_i(x)$)
as the moment when the droplet with the given $x$ has
been formed (as a droplet)
we can consider all functions of time as the
functions
of \( x \) .
Hence, we can see that the kinetic equation is reduced  to the
fact that every droplet keeps the constant value of $x$. To reconstruct the
picture
of the evolution one must establish the dependencies $t(z)$ and $\zeta(x)$.

\section{The system of the equations of condensation}

To describe the action of
the external conditions we shall introduce the value
of
the ideal supersaturation
$$
\zeta_{id} = \frac{n_{tot}}{n_{\infty}} - 1
$$

Due to solvatation of the heterogeneous centers
the ideal supersaturation falls to the value
\begin{equation} \label{4}
\Phi = \zeta_{id}  - \frac{
\sum_{i} \eta_{tot\ i}\nu_{e\ i}}{n_{\infty}}
\end{equation}
where \( \nu_{e\ i} \)
is the number of the molecules of the condensated
phase  in the equilibrium heterogeneous embryo.

One
can assume that \( \nu_{e\ i} \) can be taken
 at \( \zeta = \zeta_{*} \).

One can put approximately
$$ \Phi = \zeta_{id}$$
with a small  relative  error.  We  shall  call  $\Phi$  the  ideal
supersaturation
also.

To construct the mathematical model we must formulate some statements:
\begin{itemize}
\item
(1) The main role in the vapor consumption
(when this consumption is really essential) during the evolution
is played by the super-critical embryos, i.e. by the droplets.
\item
(2) The quasistationary approximation for the nucleation rate can be accepted
during every
period of the essential formation of the droplets on every sort of
the heterogeneous
centers (except the situations when the considered sort of the
heterogeneous centers is exhausted).
When this approximation is essential it is valid. Namely, it is not valid only
in the
situation when all the nucleus of the given sort become the centers of the
droplets. We shall call this situation as the "exhausted centers situation"
(ECS). In this very situation the quasistationary approximation isn't
necessary as far as the result of the process is obvious.
\item
(3) It is possible to change $t_*$ in the formalism of the iteration procedure
in \cite{Novosib} to the choice of $t_{**}$ which corresponds to the maximum
of the supersaturation. Really, the "correct" choice $t_*$ as the time when
the half of the total number of the droplets is already formed begins to deviate
essentially from the choice of $t_{**}$ only in ECS when the result is already
known. Certainly, the parameter in the linearization must be reconsidered
if it is necessary\footnote{Here the simplest explanation is given. The
precise analytical analysis which confirm this statement is also available.}.
\end{itemize}

The justification of the second statement uses the estimate for the times
 \( t^{s}_{i} \)
of the establishing of the stationary  state
in the near-critical region which can be found in \cite{3},
\cite{17}
(for the heterogeneous barrier  the consideration is  quite the same).
Here it is necessary to remark that,
certainly, there may exists      some
rather huge times $t^{s}_{i}$. They correspond to rather big values
of the number of the molecules in the critical embryos.
 The halfwidth of the near-critical region
\cite{3}
can be estimated by the homogeneous value at the supersaturation corresponding
to same  value of the volume of the critical embryo.
 It is proportional to    the value near
$\nu^{2/3}_{c\ i}$. Here and in
the further considerations the lower index "${c}$"
marks the values for the critical embryos.
As far as the  absorption ability is
proportional to $\nu^{2/3}_{c\ i}$
and the size of the near-critical region is proportional to $\nu^{2/3}_{c\
i}$ the
value of $t^{s}_{i}$
is proportional to $\nu^{2/3}_{c\ i}$. Practically,
the big value of the activation barrier $\Delta F_i = F_i(\nu_c)$
 (in the case when
$\nu_{c\ i}$ is  greater than the characteristic
length $\Delta x$ of the  size spectrum) means that these
kinds of the heterogeneous centers are excluded
from the kinetic process.

For the majority of the types of the heterogeneous centers
 the following approximations  for the  nucleation rate \( J_{i} \)
 are valid during the period of the essential formation of the droplets
on the centers of the given sort
\begin{equation}          \label{5}
J_{i} = J_{i}( \eta_{tot\ i} , \zeta_{*} )
\exp ( \Gamma_{i} \frac{ ( \zeta - \zeta_{*} ) }
{ \zeta_{*} } )
\frac{\eta_i}
{\eta_{tot\ i}}
\end{equation}
where
\begin{equation}                   \label{6}
\Gamma_{i} = -\zeta_{*}
\frac{d \Delta_{i} F(\zeta)}{d \zeta }  \mid_{\zeta=\zeta_{*}}
\end{equation}
The validity of these approximations is justified
for the heterogeneous embryos with
the monotonous interaction between the nucleus and the molecules
of the condensated phase
which is  weaker or equal than the reciprocal  to the space distance.

Let \( f_{*\ i} \) be the amplitude value of the
size distribution  of the heterogeneously formed droplets
 measured in the units of \( n_{\infty} \).
As far as the supersaturation \( \zeta_{*} \) and the number of
the heterogeneous
centers \( \eta_{tot\ i} \) are the already
known values the stationary rate of nucleation $J$ and
the stationary distribution $f$
 can be easily calculated by some formulas derived in \cite{18}:
\begin{equation} \label{7}
\frac{J_i \tau}{\alpha \zeta n_{\infty}} =
f=\frac{ W^{+}_{c} \exp(-\Delta_i F) \tau}
{ n_{\infty} \pi^{1/2} \Delta_{e\ i} \nu  \Delta_{c\ i} \nu  \zeta  \alpha }
\eta_{i}
\end{equation}
where $W^{+}$ is the number of the molecules absorbed by  the embryo  in
the unit
of time, $\Delta_{e} \nu$  is the width of the equilibrium distribution
$$
\Delta_{e\ i} \nu = \sum_{\nu=1}^{\nu=(\nu_{c}+\nu_{e})/2}\exp(-F(\nu))
$$
and $\Delta_{c\ i} \nu$ is the halfwidth of the near-critical region
$$
\Delta_{c} \nu =
\frac{2^{1/2}}{\mid \frac{\partial^{2} F }{\partial
\nu^{2}}\mid^{1/2}_{\nu=\nu_{c}}}
$$

We shall mark by \( n_{\infty} g_{i} \) the total number of the vapor
molecules in the   droplets
formed on the centers of the  sort "$i$".
To simplify the formulas we shall use $$ \theta_{i} =
\frac{\eta_{i}}{\eta_{tot\ i } } $$

Let us choose for the simplicity the concrete conditions of the constant
volume and pressure of the system. This allows to
cancel some unimportant parameters
and  to exclude them from formulas.

Using the conservation laws for the heterogeneous centers
and for the molecules of the condensated
substance   we obtain for \( g_{i},  \theta_{i} \) the following
equations
\begin{equation}\label{8}
g_{i} = f_{*\ i}  \int_{-\infty}^{z_i} (z_i-x_i)^{3}
\exp ( \Gamma_i \frac{ \Phi(x) - \zeta_{*\ i}}{\zeta_{*\ i}})
\exp ( -\Gamma_{i} \frac{ \sum_{j}g_{j}  }
{ \zeta_{*\ i} } )
\theta_{i} dx_i
\equiv
G_{i}(\sum_{j}g_{j}, \theta_{i} )
\end{equation}
\begin{equation}\label{9}
\theta_{i} = \exp ( - f_{*\ i} \frac{n_{\infty}}{\eta_{tot\ i}}
\int_{-\infty}^{z_i}
\exp ( \Gamma_i \frac{ \Phi(x) - \zeta_{*\ i}}{\zeta_{*\ i}})
\exp ( - \Gamma_{i} \frac{ \sum_{j}g_{j}  }
{ \zeta_{*\ i} } ) dx_i )
\equiv
S_i( \sum_{j}g_{j})
\end{equation}
where $f_{*\ i} = J_{i}(\eta_{tot\ i},\zeta_{*\ i})\tau /
\zeta_{*\ i} \alpha n_{\infty}$.

These equations form the closed system of the equations of the condensation
kinetics. This system will be the subject of our investigation.

As far as we measure the accuracy of the theory in the terms of
the error in the number of the droplets
 we  define these values as the following ones:
\begin{equation} \label{10}
N_{i} = \eta_{tot\ i} ( 1 - \theta_{i}(z)) \equiv Q_{i}(\theta_{i})
\end{equation}
The size spectrum  can be found as the following one
\begin{equation} \label{11}
f_{i}=
f_{*\ i}
\exp ( -\Gamma_{i} \frac{ \sum_{j}g_{j}  }
{ \zeta_{i\ *} } )
\exp ( \Gamma_i \frac{ \Phi(x) - \zeta_{*\ i}}{\zeta_{*\ i}})
\theta_{i}
\end{equation}

Note  that  system  (\ref{8})  -  (\ref{9})  is  rather   more
complex
than in the case of decay. The specific
 feature is not only the presence
of the factor $
\exp ( \Gamma_i \frac{ \Phi(x) - \zeta_{*\ i}}{\zeta_{*\ i}})   $
but also the necessity of
the introduction of the separate scales for $z$ and $x$.
It means that the periods of the
essential formation of the droplets on the centers
of a different nature don't coincide or have any common regions. Such a situation
can be guaranteed by the following inequalities
\begin{equation} \label{ier}
\frac{f_{*\ i}}{f_{*_j}} \ll 1
\end{equation}
or
\begin{equation} \label{ier1}
\frac{f_{*\ i}}{f_{*_j}} \gg 1
\end{equation}
for  all arbitrary pairs of the sorts of the centers $i$ and $j$.
The situation
\begin{equation} \label{noier}
\frac{f_{*\ i}}{f_{*_j}} \sim 1
\end{equation}
will  be  at  first  excluded   from   consideration.   Later   the
generalization
will be presented.

Note that  inequalities  (\ref{ier})  -  (\ref{noier})  are  rather
unsensitive
to the change of the supersaturation. Note that
$$
f_* \sim \exp(-F_c + F_e)
\sim \exp(-F_{c\ hom} + F_e)
$$
where the free energy $F$ is counted in the "absolute" scale
with the zero point of the energy corresponding
to the standard homogeneous zero point $F_{hom}(\nu=1) = 0$ and
$F_{j\ e}$ are the energies of solvatation,
So, we can write
\begin{equation} \label{simier}
\frac{f_{*\ i}}{f_{*_j}} \sim \frac{\eta_{i\ tot} \exp(+F_{i e})}
 {\eta_{j\ tot} \exp(+F_{j e})}
\end{equation}
Hence, we can see that
when the free energy of solvatation is unsensitive to the supersaturation (as
it occurs really in the nature) then the r.h.s. of the last equation doesn't
depend on the supersaturation.

So, the set of the heterogeneous centers $\{ \eta_{i\ tot} \}$ initiates the
set of the characteristic times $\{t_{*i} \}$ and the characteristic
supersaturations
$ \{ \zeta_{i*} \}$.
Suppose that $ \zeta_{i\ *} > \zeta_{*\ j}$ if $i>j$.

\section{External supersaturation}

The analysis of the last section simplifies the treatment, but isn't
absolutely necessary. The more rigorous recipe is to solve the separate problems
\begin{equation}\label{8a}
g_{i} = f_{*\ i}  \int_{-\infty}^{z_i} (z_i-x_i)^{3}
\exp ( \Gamma_i \frac{ \Phi(x) - \zeta_{*\ i}}{\zeta_{*\ i}})
\exp ( -\Gamma_{i} \frac{ g_{i}  }
{ \zeta_{*\ i} } )
\theta_{i} dx_i
\equiv
G_{i}(g_{i}, \theta_{i} )
\end{equation}
\begin{equation}\label{9a}
\theta_{i} = \exp ( - f_{*\ i} \frac{n_{\infty}}{\eta_{tot\ i}}
\int_{-\infty}^{z_i}
\exp ( \Gamma_i \frac{ \Phi(x) - \zeta_{*\ i}}{\zeta_{*\ i}})
\exp ( - \Gamma_{i} \frac{ g_{i}  }
{ \zeta_{*\ i} } ) dx_i )
\equiv
S_i( g_{i})
\end{equation}

This solution can be obtained by the methods from \cite{Novosib} and we needn't
to describe it here.
As a result we get the set $\{\zeta_{*\ i}\}$. The solution is valid only for
$i=1$ ($\zeta_{*\ 1} < \zeta_{*\ i \neq 1} $). All other solutions are got
only to
construct the set $\{ \zeta_{*\ i} \}$. Now they are useless.

 The further condensation can't be described by  system (\ref{8a}) -
(\ref{9a})
because the droplets formed on the centers of the first sort consumes
the vapor. This system must be substituted by
\begin{equation}\label{8ab}
g_{i} = f_{*\ i}  \int_{-\infty}^{z_i} (z_i-x_i)^{3}
\exp ( \Gamma_i \frac{ \Phi(x) - \zeta_{*\ i}}{\zeta_{*\ i}})
\exp ( -\Gamma_{i} \frac{ g_{i} + g_{1}  }
{ \zeta_{*\ i} } )
\theta_{i} dx_i
\end{equation}
\begin{equation}\label{9ab}
\theta_{i} = \exp ( - f_{*\ i} \frac{n_{\infty}}{\eta_{tot\ i}}
\int_{-\infty}^{z_i}
\exp ( \Gamma_i \frac{ \Phi(x) - \zeta_{*\ i}}{\zeta_{*\ i}})
\exp ( - \Gamma_{i} \frac{ g_{i} + g_{1} }
{ \zeta_{*\ i} } ) dx_i )
\end{equation}
Here $i \neq 1$. In order to simplify this system it is necessary to take
into account the following statement
\begin{itemize}
\item
The size spectrum of the droplets of those sorts for which the formation
of the droplets has been already described can be treated as the set of
the monodisperse
spectrums with some coordinates (depended on the sort of the centers).
\end{itemize}
The solution of the system of the condensation equations  was necessary to get
the number of the droplets and the coordinate of the spectrum.

Then the system of the condensation equations can be reformulated in the
 following way
\begin{equation}\label{8b}
g_{i} = f_{*\ i}  \int_{-\infty}^{z_i} (z_i-x_i)^{3}
\exp ( \Gamma_i \frac{ \Phi(x) - \zeta_{*\ i}}{\zeta_{*\ i}})
\exp ( -\Gamma_{i} \frac{ g_{i} + \frac{N_{tot\ 1}}{n_{\infty}} z^3_{1\ i}  }
{ \zeta_{*\ i} } )
\theta_{i} dx_i
\end{equation}
\begin{equation}\label{9b}
\theta_{i} = \exp ( - f_{*\ i} \frac{n_{\infty}}{\eta_{tot\ i}}
\int_{-\infty}^{z_i}
\exp ( \Gamma_i \frac{ \Phi(x) - \zeta_{*\ i}}{\zeta_{*\ i}})
\exp ( - \Gamma_{i} \frac{ g_{i} + \frac{N_{tot\ 1}}{n_{\infty}} z^3_{1\ i} }
{ \zeta_{*\ i} } ) dx_i )
\end{equation}
Here $i \neq 1$, $N_{tot\ 1}$ is the
total number of the droplets formed on the
heterogeneous centers of the first sort and $z_{1\ i}$ is the coordinate
of spectrum of the sizes of the droplets formed on the heterogeneous centers
of the first sort.

Actually $$N_{i\ tot} \sim \eta_{i\ tot}$$
If the last equality isn't valid it means that the supersaturation is
going to attain soon
 the maximum  and then to fall.  So,  no  other  formation  of
the droplets can
be observed.

One must solve the last system for all sorts of the heterogeneous centers
and construct the sequence $\{ \zeta_{*\ i} \}_{i \geq 2}$. Ordinary the
sequence
will be the old one, but the concrete values are changed.

Again we can use only the lowest $\zeta_{*\ i}$, i.e. the second sort. Then
we must repeat this procedure again and again. The question  how to solve
these systems will be discussed later.

After condensation on $i_0$ sorts of the heterogeneous centers
has been  already
described and we are going to describe condensation on the centers
of the sort $i$ the following system must be solved
\begin{equation}\label{8c}
g_{i} = f_{*\ i}  \int_{-\infty}^{z_i} (z_i-x_i)^{3}
\exp ( \Gamma_i \frac{ \Phi(x) - \zeta_{*\ i}}{\zeta_{*\ i}})
\exp ( -\Gamma_{i} \frac{ g_{i} + \sum_{j \leq i_0} \frac{N_{tot\
j}}{n_{\infty}}
 z^3_{j\ i}  }
{ \zeta_{*\ i} } )
\theta_{i} dx_i
\end{equation}
\begin{equation}\label{9c}
\theta_{i} = \exp ( - f_{*\ i} \frac{n_{\infty}}{\eta_{tot\ i}}
\int_{-\infty}^{z_i}
\exp ( \Gamma_i \frac{ \Phi(x) - \zeta_{*\ i}}{\zeta_{*\ i}})
\exp ( - \Gamma_{i} \frac{ g_{i} + \sum_{j \leq i_0}
\frac{N_{tot\ j}}{n_{\infty}} z^3_{j\ i} }
{ \zeta_{*\ i} } ) dx_i )
\end{equation}
Here $i > i_0$, $N_{tot\ j}$ is the total number of the droplets formed on
the centers of the sort $j$ (it is near to $\eta_{j \ tot}$),
$z_{j\ i}$ is the coordinate of these droplets.

System (\ref{8c}) - (\ref{9c}) is rather complex because the
values $z_{j\ i}$ are unknown ones. The next statement simplifies the
problem and reduces it to the two more simple problems:
\begin{itemize}
\item System (\ref{8c}) - (\ref{9c}) can be with a high accuracy reduced
to the following system
\begin{equation}\label{8d}
g_{i} = f_{*\ i}  \int_{-\infty}^{z_i} (z_i-x_i)^{3}
\exp ( \Gamma_i \frac{ \Omega(x) - \Omega_{*\ i}}{\Omega_{*\ i}})
\exp ( -\Gamma_{i} \frac{ g_{i} }
{ \Omega_{*\ i} } )
\theta_{i} dx_i
\end{equation}
\begin{equation}\label{9d}
\theta_{i} = \exp ( - f_{*\ i} \frac{n_{\infty}}{\eta_{tot\ i}}
\int_{-\infty}^{z_i}
\exp ( \Gamma_i \frac{ \Omega(x) - \Omega_{*\ i}}{\Omega_{*\ i}})
\exp ( - \Gamma_{i} \frac{ g_{i}  }
{ \Omega_{*\ i} } ) dx_i )
\end{equation}
where the  function $\Omega$ satisfies the following equation
\begin{equation} \label{Om}
\Phi = \Omega + \sum_{j \leq i_0} \frac{N_{j\ tot}}{n_{\infty}} (z+\delta z_j)^3
\end{equation}
\begin{equation} \label{Om1}
\Omega =
\frac{\tau}{\alpha} \frac{dz}{dt}
\end{equation}
and $\delta z_j$ are some fixed and already known distances between
the monodisperse
spectrums.
\end{itemize}

Here the following statement is also taken into account
\begin{itemize}
\item
The following approximation during the period of the intensive formation of
the droplets on the heterogeneous centers of the given sort (PIFDGS) is valid
for the nucleation rates \( J_{i} \)
\begin{equation}          \label{5a}
J_{i} = J_{i}( \eta_{tot\ i} , \Omega_{*} )
\exp ( \Gamma_{i} \frac{ ( \zeta - \Omega_{*} ) }
{ \Omega_{*} } )
\frac{\eta_i}
{\eta_{tot\ i}}
\end{equation}
where
\begin{equation}                   \label{6a}
\Gamma_{i} = -\Omega_{*}
\frac{d \Delta_{i} F(\zeta)}{d \zeta }  \mid_{\zeta=\Omega_{*}}
\end{equation}
and the index "*" corresponds to $t_{*\ i}$.
\end{itemize}

The validity of
the last system of equations  is based  on  the simple fact that the rate of
the vapor consumption
is violated by the relatively essential deviation of the supersaturation
and the rate of nucleation is violated by the relatively small violation of
the supersaturation corresponding to the violation of $\frac{\Gamma_i}{\zeta}$.

We shall call $\Omega$ the external supersaturation. Our next task is to determine
it.

\section{Iterations for the external supersaturation}

 System (\ref{Om}) - (\ref{Om1}) forms the ordinary differential equation
(of the Abel's type) and doesn't allow the analytical solution.
The initial condition is the following
$$ z \mid_{t=t_{init}} = 0$$
where $t_{init}$ is the time of formation of the last known peak of the
droplets
distribution.

Meanwhile one can construct  the iterations on the base of the small parameter
\begin{equation}
\delta =
\frac{ \sum_{j \leq i_0} N_{j\ tot} (z+\delta z_j)^3 }{\Phi n_{\infty}}
\end{equation}
              We suppose that
$$ \frac{d^2 \Phi}{dt^2}
\leq
0
$$
and can announce the following statements
\begin{itemize}
\item
The iterations defined according to
$$
\Phi = \frac{\tau}{\alpha} \frac{dz_{(k+1)}}{dt}
+ \sum_{j \leq i_0} \frac{N_{j\ tot}}{n_{\infty}} (z_{(k)}+\delta z_j)^3
$$
converge to a unique solution (an ordinary property of
a differential  equation)\footnote{We shall put for simplicity $z_{(0)}=0$.}
\item
We need to investigate the region where
$$ \frac{d\Omega}{dt} \geq 0$$
When $\frac{d\Omega}{dt} = 0$ at the final of the period under
the consideration
we have
$$ \frac{d | max \Omega_{(k)} -
max \Omega_{(k+1)} | }{d(\frac{d^2\Phi}{dt^2})} \geq 0 $$
Actually we can remove $t_{init}$ to the time of formation of the first
peak and consider all $\delta z_j$ as the positive values.
Then
$$ \frac{d | max \Omega_{(k)} - max \Omega_{(k+1)} | }{d \delta z_j} \leq 0 $$
 The previous  inequality
in such a reformulation can
be also justified.
Another inequality can be also taken into account
$$ \frac{d | max \Omega_{(k)} - max \Omega_{(k+1)} | }{dt_{init}} \geq 0 $$
Certainly, $t_{init}$ can not be put earlier than the moment when
$\Phi = 0$.
\end{itemize}

So in the worst situation the form of the differential equation is the following:
\begin{equation} \label{linear}
A t = \frac{\tau}{\alpha} \frac{dz}{dt}
+   \frac{N_{tot}}{n_{\infty}} z^3
\end{equation}
where $A$ is some constant and $N_{tot}$ is some value.

If $\Phi$ is the polinom then all iterations are calculated analytically
and have the form of polinoms. Their convergence is rather high. The value
of  a relative error in the maximum of $\Omega$
$$                    \Delta_{i} \leq \frac{ | max_{t} \Omega_{(i)} -
 max_{t} \Omega|} { max_{t} \Omega } $$ is less than
$$                    \Delta_{i\ i+1} \sim \frac{ | max_{t} \Omega_{(i)} -
 max_{t} \Omega_{(i+1)}|} { max_{t} \Omega_{(i+1)}} $$
and can be easily calculated on the base of (\ref{linear}).
As a result one can get
$\Delta_2 < 4.2*10^{-2}$, $\Delta_3 < 8.68*10^{-3}$.

Note that the requirement for the choice of the number of iterations is the
following
$$ \Delta_k \leq \frac{1}{\Gamma_i}$$
Really, $\Gamma_i$ is the big parameter of the theory and goes to infinity,
but actually $\Gamma_i \sim 50$ which is necessary for the intensive
formation of the droplets ($f_* \sim \exp( - const \Gamma_i^{2/3})$).
So, we need to calculate only the few first iterations.

Figure 1 shows the form of the iterations. Note that the evident renormalization
can lead to the differential equation without parameters
\begin{equation} \label{linear2}
t = \frac{dz}{dt}
+    z^3
\end{equation}
with the previous initial condition.

The forms of the iterations $f_1$ -  $f_4$
are drawn in Figure 1 where the inex points the number of the iteration.
 Note, that only the
behavior until the maximum is interesting for us. It can be seen that
$f_3$ can not already be separated from $f_4$.

One can also introduce another  approximation which is more accurate.
This is the approximation of the ideal surface.
In this approximation the term $z^3$ is
split between the surface term $z^2$ and the linear term $z$.
Return to the previous scales of the variables. The iterations
are constructed by the following way
$$
\Phi = \frac{\tau}{\alpha} \frac{dz_{(k+1)}}{dt}
+ \sum_{j \leq i_0} \frac{N_{j\ tot}}{n_{\infty}} (z_{(k)}+\delta z_j)^2
 (z_{(k+1)}+\delta z_j)
$$
The rate of convergence of such iterations is higher than of the old ones,
but there is no opportunity to calculate iterations analytically.
Hence, this way of the approximation can be effectively used only at the last
step.

Now we are going to solve  system (\ref{8d}) - (\ref{9d}). This system
is more complex than the system already considered in \cite{Novosib} because
the behavior of $\Omega$ during PIFDGS
 is essentially nonlinear. One can not linearize
$\Omega$ as the function of time and $x$ during PIFDGS.
But some important conclusions about
the behavior of $\Omega$ can be made.
Consider an arbitrary sort of the heterogeneous centers $i$ on which the droplets
have been  already formed. Introduce the characteristic width $ \Delta x$
of the peak which
is going to be formed.
Certainly,
\begin{equation} \label{jkl}
 \Delta x \ll z_i
\end{equation}
The value of $g_i$ near $z \sim 0$ allows to give the estimate
$$g_{i} \sim \frac{N_{i\ tot}}{n_{\infty}} z_i^3$$
     We can also give the analogous estimates for the derivatives
$$\frac{d g_{i}}{d z} \sim \frac{N_{i\ tot}}{n_{\infty}} z_i^2$$
$$\frac{d^2 g_{i}}{dz^2} \sim \frac{N_{i\ tot}}{n_{\infty}} z_i$$
$$\frac{d^3 g_{i}}{dz^3} \sim \frac{N_{i\ tot}}{n_{\infty}} $$
           The actions of this terms on $\Omega$ are given by the following
terms          (i.e. the corresponding terms in the Tailor's series)
$$\frac{d g_{i}}{d z} \rightarrow \frac{N_{i\ tot}}{n_{\infty}} z_i^2
\Delta x$$
$$\frac{d^2 g_{i}}{dz^2} \rightarrow \frac{N_{i\ tot}}{n_{\infty}} z_i
(\Delta x)^2$$
$$\frac{d^3 g_{i}}{dz^3} \rightarrow \frac{N_{i\ tot}}{n_{\infty}} (\Delta
x)^3$$
                                    Due to (\ref{jkl}) one can see that
$$       \frac{N_{i\ tot}}{n_{\infty}} z_i^2
\Delta x
\gg
\frac{N_{i\ tot}}{n_{\infty}} z_i
(\Delta x)^2
\gg
 \frac{N_{i\ tot}}{n_{\infty}} (\Delta x)^3
$$
The action of
$\frac{d g_{i}}{d t}$ can be cancelled by $\frac{d\Phi}{dt}$ because
they have the different signs. So the action of $\frac{d^2 g_{i}}{dt^2}$ must be
taken
into account. The action of $\frac{d^3 g_{i}}{dt^3}$ can be neglected.

As the result we justify the following approximation
for the behavior of $\Omega$ during
                                 PIFDGS:
\begin{equation}
\frac{\Gamma}{\Phi_*} \Omega = cx + lx^2 + const
\end{equation}
with the two known constants.

The characteristic length of formation of the spectrum by means of the
vapor
consumption
by the droplets in this very peak is given by the following formula
$$
\Delta x \sim (\frac{\Omega_{i\ *}}{\Gamma_i f_{*\ i}})^{1/4}
$$
The halfwidth of the spectrum initiated by the external supersaturation is given
by
$$
x_p = (\frac{2 \Omega_{max}}{\Gamma_i | \frac{d^2
\Omega}{dx^2}\mid_{\Omega = \Omega_{max}} | })^{1/2}
$$
when formation of the droplets occurs near the maximum of $\Omega$.

When $\Delta_x \ll x_p$ we can use the standard iteration procedure described
in \cite{Novosib}. When $\Delta x \gg x_p$ or $\Delta x \sim x_p$ we must
use the modified method of the steepens descent \cite{Deponall}
which is described in the next section
(it covers all situations).
The case $\Delta x \gg x_p$ is extremely simple ($\zeta$ coincides with
$\Omega$).

\section{Modified method of the steepens descent}

In this section we shall forget about the several types of the heterogeneous
centers and about the several peaks in the spectrum of the droplets sizes.
We shall consider the model situation when we have the nonlinear
external source of the supersaturation of the square form and the process
of the homogeneous (or the heterogeneous) condensation.  We shall start with the
homogeneous condensation and then the generalization on the heterogeneous
case will be given. According to the results of the previous section this
model covers all possible situations.

\subsection{Homogeneous condensation}

In order to obtain the form of the  function $\zeta(x)$ the
balance equation must be introduced. For the homogeneous condensation
one can obtain       the following balance equation
\begin{equation} \label{h1}
\Phi = \zeta + g
\ \ \ \ \ \ \
g(z) = \frac{n_{\infty\ *} V_{*}}{n_{\infty} V}
\int_{-\infty}^{z} dx (z-x)^{3} f(x)
\end{equation}
The value of $t_*$ can be chosen as the moment of the maximum of the
supersaturation.

The term $ \frac{n_{\infty\ *} V_{*}}{n_{\infty} V} $ is approximately
equal to $1$
and can  be missed.

The value of the
ideal supersaturation  can be presented
in the following form
\begin{equation}\label{h2}
\Phi = \Phi_{*} + \Phi_{*} ( cx + lx^2)/\Gamma
\end{equation}
with the two parameters
\begin{equation}\label{h}
c=\frac{\Gamma}{\Phi_{*}} \frac{d\Phi}{dx}\mid_{x=0} \ \ \
l=\frac{\Gamma}{2 \Phi_{*}} \frac{d^2\Phi}{dx^2}\mid_{x=0}
\end{equation}
as it was done in \cite{small}.
The negative value of $l$ corresponds to the nonlinear character of
the external
conditions.
The form of the ideal supersaturation  (\ref{h2}) can be
treated as the Tailor's
series cut of on the second term.
The result of \cite{small} states the validity of (\ref{h2})
in the case when the nonlinearity of
the external
source
is induced by the vapor consumption of
the already formed spectrums.

The two assumptions justified analytically can be introduced. The
first one establishes that
\begin{itemize}
\item
the leading role in the vapor
consumption belongs to the super-critical
embryos, i.e. to the  droplets.
\end{itemize}

The second assumption states
\begin{itemize}
\item
the quasistationary state of the embryos
in the near-critical region.
\end{itemize}
It must be valid during the period of
the essential formation of the droplets. Let $t_{\zeta}$ be the characteristic
time of the variation of the stationary state in the near-critical region and
let $t_{s}$  be the time of the relaxation to the stationary state in
the near-critical region.
Then the required assumption can be written as the following one
\begin{equation}
t_{\zeta} \gg t_{s}
\end{equation}

These assumptions will be valid in all situations of
the homogeneous case and in
the heterogeneous case except the situation
when almost all heterogeneous centers are exhausted at the final of
the period of the essential formation.
In this situation the result of
the period of the essential formation
of the droplets is obvious: the number of
the droplets coincides with the total number of the heterogeneous centers.
The form of the spectrum is unessential during the period of
the essential formation of the droplets
and the spectrum  is the monodisperse one after the supersaturation
begins to fall.

Note that the second statement may be not valid in some situations when
the supersaturation doesn't depend of the process of the vapor consumption
by the droplets (it is negligible).
 As it can be shown in these situations the effects of
the nonstationarity aren't important for  the results of the process\footnote{This
case requires the special consideration which can be done and shows that
the effects of the nonstationarity are cancelled.}.

Now let us  analyze the behavior  of  $g(z)$.  At  first  we  shall
analyze
it only qualitatively. We shall extract the three regions. Let us put $l \sim
0 $. At first we shall define the period of
the creation  of the main  consumers of the vapor during the
period of the essential formation of the droplets. Then the behavior of
$g$ in the first iteration can be seen from
\begin{equation}\label{h4}
g(z) \sim \exp(z) \int_{0}^{\infty} y^{3} \exp(-y)dy
\end{equation}
One can note the following important fact:
\begin{itemize}
\item
The subintegral function in (\ref{h4}) is sufficient only when
\begin{equation}\label{h4'}
1 \sim  y_{min} \leq  y \leq y_{max} \sim 8\ \ \ \ y= c\rho
\end{equation}
This region is the region of the creation of the main consumers of the vapor
during the period of the essential formation of the droplets. It is
moving
in time along the $\rho$-axis (and the $t$-axis) with the velocity of the
droplets
growth\footnote{Only until the end of the process of the creation of the
droplets because the deviation of the supersaturation from the ideal one
essentially transforms the definitions of the boundaries of this region}.
When the last inequalities aren't valid the subintegral function
is negligible. The period which corresponds to  (\ref{h4'})  plays the main
role
in formation of the spectrum.
\end{itemize}
Now we shall extract the  other regions
\begin{itemize}
\item
During  the period of the essential formation of the droplets one can extract the
"initial" region according to
\begin{equation}\label{hd}
-9 \leq cz \leq c z_{b}
\end{equation}
The value of $z_b$ must be established with the help of the additional condition
\begin{equation}\label{hdd}
\frac{f(x) - f_{1}(x) }{ f_{1}(x)} \ll 1
\end{equation}
where $f$ is the size distribution of the droplets and
$f_{1}$ is the imaginary distribution formed at the supersaturation $\Phi$
instead of $\zeta$.
The values corresponding to this region will be marked by the subscript "$in$".

\item
The estimates for $g(z)$ give the above estimate for the duration of the
period of the essential formation of the droplets. The period of the intensive
 formation of
the droplets finishes when $cz \sim 1$. It starts when $cz \sim -1$.
One can define the peak  of the period of the intensive formation
of the droplets according to
\begin{equation}\label{hd2}
- c^{-1} \leq z \leq c^{-1}
\end{equation}
The successful solution of the system of the condensation equations on the
second step of iterations  shows  that  the
subintegral
function lies essentially inside the initial period during the intensive formation
of the droplets. Then one can conclude that\footnote{and even
$$
- y_{min} + 1 \sim c z_b
$$
}
\begin{equation}\label{hd1}
- y_{min} \leq cz_{b}
\end{equation}
If $l \leq 0$ (namely this case must be considered), then
the hierarchy becomes even more strong.
\end{itemize}

So, the region of the creation of the main consumers of the vapor during
the period of the essential formation ofthe droplets belongs to the initial
region.

\subsection{Approximation of the spectrum}

One need to construct some approximation for $f(x)$ at
 the period of the essential formation of the droplets. The problem
can be reduced
to  the suitable expression for $\zeta - \zeta_{*}$ in the
standard exponential approximation
\begin{equation}\label{hd3}
f(x) = f_{*} \exp(\frac{\Gamma (\zeta - \zeta_{*})}{\Phi_{*}})
\exp(\frac{\Gamma (\zeta_{*} - \Phi_{*})}{\Phi_{*}})
\end{equation}

One can decompose $\zeta - \zeta_{*}$ into the Tailor's series  in  the
neighborhood
of $z=0$. The expressions for the derivatives are the following ones:
\begin{equation}\label{hd4}
\frac{d\zeta}{dz} = \frac{d\Phi}{dz} - 3 \int_{-\infty}^{z}
f(x) (z-x)^2 dx
\end{equation}

\begin{equation}\label{hd5}
\frac{d^2\zeta}{dz^2} = \frac{d^2\Phi}{dz^2} - 6 \int_{-\infty}^{z}
f(x) (z-x) dx
\end{equation}

\begin{equation}\label{h10}
\frac{d^3\zeta}{dz^3} = \frac{d^3\Phi}{dz^3} - 6 \int_{-\infty}^{z}
f(x)  dx
\end{equation}

\begin{equation}\label{h11}
\frac{d^n\zeta}{dz^n} = \frac{d^n\Phi}{dz^n} - 6 \frac{d^{n-4} f(z)}{dz^{n-4}}
\ \ \ \ \ \ \ n \geq 4
\end{equation}

The value $d^3 g / d z^3 $ is proportional to the number of the already
formed droplets.
This value is the main result of the process of condensation and the
subject of interest of the  theoretical description. According to the
iteration procedure \cite{Novosib} one can note that

\begin{itemize}
\item
        1). At the first step of the iteration procedure
the expression for the total number of the droplets isn't correct at all.
The relative error goes to $\infty$.
\item
        2). The second step of the iteration procedure is the final
one and it gives           the
almost precise expressions for all main parameters of the process of
condensation.
  \end{itemize}

Hence, one can introduce the procedure without
the account of (\ref{h10}), (\ref{h11}).
 Then one can can restrict  the Tailor's
series by
the first two  terms. This restriction leads to the following approximation
\begin{equation}\label{h12}
f(x) = f_{*} \exp(\Gamma \frac{ d^2 \zeta}{2dz^2}\mid_{z=0}
\frac{x^2} {\Phi_{*}})
 \exp(\frac{\Gamma (\zeta_* - \Phi_{*})}{\Phi_{*}})
=
f_{m}
\exp(\Gamma \frac{d^2 \zeta}{2 d z^2} |_{z=0} \frac{x^2}{\Phi_*} )
\end{equation}
This approximation is valid only inside the peak of the period of the
essential formation of the droplets. It fails outside this region.
The standard  method of the steepens descent  spreads
 approximation (\ref{h12}) on the
whole period of the essential formation of the droplets
(including the "initial" region and the region of the creation of the
main consumers of the vapor during the period of the essential formation
of the droplets) which leads to the essential
relative error.
Moreover,   approximation (\ref{h12}) isn't necessary because in the
initial
region the correct expression for the spectrum has been already established
\begin{equation}\label{h13}
f(x) = f_{*} \exp(\Gamma \frac{\Phi(x) - \Phi_{*}}{\Phi_{*}})
\end{equation}

\subsection{Equations on spectrum parameters}

 To obtain the system of equations on the parameters of the spectrum one can
differentiate the balance equation for the condensated substance at $z=0$. Then
the
following equations appear:
\begin{equation}\label{hd6}
\Phi(0) = g_{in}(0) + g_{ex}(0) +\zeta(0)
\end{equation}

\begin{equation}\label{hd7}
\frac{d\Phi}{dz}\mid_{z=0}  =\{ \frac{dg_{in}}{dz} +\frac{d g_{ex}}{dz} +
\frac{d\zeta}{dz} \} \mid_{z=0}
\end{equation}

\begin{equation}\label{hd8}
\frac{d^2\Phi}{dz^2}\mid_{z=0}  =\{ \frac{d^2g_{in}}{dz^2}
+\frac{d^2 g_{ex}}{dz^2} +
\frac{d^2\zeta}{dz^2}  \} \mid_{z=0}
\end{equation}
The subscripts "$in$" and "$ex$" mark the part of $g$ referred to the
droplets formed inside the "initial" region and  outside it. Equations
(\ref{h13}) and (\ref{h2}) lead to
\begin{equation}\label{hd9}
g_{in} = \sum_{i=0}^{3} z^{3-i} \alpha_{i}
\end{equation}

\begin{equation}\label{hd10}
\frac{d g_{in}}{dz} = \sum_{i=0}^{2} z^{2-i} (3-i) \alpha_{i}
\end{equation}

\begin{equation}\label{hd11}
\frac{d^2 g_{in}}{dz^2} = \sum_{i=0}^{1} z^{1-i} (3-i)(2-i) \alpha_{i}
\end{equation}

In these equations the values $\alpha_{i}$ are defined according to
\begin{equation}\label{hd12}
\alpha_{i} = (-1)^{i} \frac{3!}{i!(3-i)!} f_{*} \int_{-\infty}^{z_{b}}
x^{i} \exp(cx+lx^2) dx
\end{equation}
To obtain the algebraic system of equations one can substitute the integral
term
in the last expression by the one of the Boyd's estimates
\begin{equation}\label{Boyd}
\frac{\pi/2}{\sqrt{z^2+\pi}+(\pi-1) z} \leq \exp(z^2)
\int_{z}^{\infty} \exp(-t^2) dt \leq
\frac{\pi/2}{\sqrt{(\pi-2)z^2+\pi}+ 2 z}
\ \ \ z>0
\end{equation}

Analogous procedure can be proposed for the values $g_{ex}, \frac{dg_{ex}}{dz},
 \frac{d^2g_{ex}}{dz^2}$. According to (\ref{h12}) the expressions for
$g_{ex}(0), dg_{ex}/dz\mid_{z=0}, d^2 g_{ex}/dz^2 \mid_{z=0}$
have the following system
\begin{equation}\label{hd13}
g_{ex}(0) = f_{m} \int_{z_{b}}^{0} \exp(-(\frac{x}{x_{p}})^2) x^3 dx =
f_{m} x_{p}^4 \lambda_{3}
\end{equation}

\begin{equation}\label{hf1}
\frac{dg_{ex}}{dz} \mid_{z=0}= 3 f_{m} \int_{z_{b}}^{0}
\exp(-(\frac{x}{x_{p}})^2) x^2 dx =
3 f_{m} x_{p}^3 \lambda_{2}
\end{equation}

\begin{equation}\label{hd14}
\frac{d^2g_{ex}}{dz^2} \mid_{z=0}= 6 f_{m}
\int_{z_{b}}^{0} \exp(-(\frac{x}{x_{p}})^2) x dx =
6 f_{m} x_{p}^2 \lambda_{1}
\end{equation}
where

\begin{equation}\label{hd15}
\lambda_{i} = \int_{\xi_{0}}^0 \xi^i \exp(-\xi^2) d\xi \ \ \
\xi_{0} = \frac{z_{b}}{x_{p}}
\end{equation}
and $x_p$ is some parameter like the characteristic halfwidth.

As the result one can see the system of the three algebraic equations which can
be solved by the standard numerical methods. The solution of this equation gives
all main characteristics of the process.

\subsection{Linear source}

In the case $l=0$ the final expressions become more simple. For $z_{b}$ one
can take here the value $-x_{p}$. Then
one can obtain
\begin{equation}\label{h111}
\alpha_{i} = f_{*} \frac{3!(-1)^i}{i!(3-i)!} c^{-i-1} \sum_{j=0}^{i}
\frac{i!}{(i-j)!} (-1)^{j} (-cx_{p})^{i-j} \exp(-cx_{p})
\end{equation}
This induces the following expressions

\begin{equation}\label{hd16}
g_{in}(0) = \alpha_{3} =  - f_{*} c^{-4} \sum_{j=0}^{3}
\frac{3!}{(3-j)!} (-1)^{j} (-cx_{p})^{3-j}
\exp(-cx_p)
\end{equation}

\begin{equation}\label{h112bis}
\frac{dg_{in}}{dz}\mid_{z=0} = \alpha_{2}
 =  f_{*} 3 c^{-3} \sum_{j=0}^{2}
 (-1)^{j} \frac{2!}{(2-i)!} (-cx_{p})^{2-j}
\exp(-cx_p)
\end{equation}
\begin{equation}\label{h112}
\frac{d^2g_{in}}{dz^2}\mid_{z=0} =2 \alpha_{1}
 = - f_{*} 6 c^{-2} \sum_{j=0}^{1}
 (-1)^{j} (-cx_{p})^{1-j} \exp(-c x_p)
\end{equation}
The values of $\lambda_{i}$ become the universal constants

\begin{equation}\label{h113}
\lambda_{i} = \int_{-1}^{0} \exp(-x^2) x^i dx
\end{equation}

Introduction of the connection between $\frac{d^2\zeta}{dz^2}$ and
$x_{p}$
\begin{equation}\label{h114}
\mid \frac{d^2\zeta}{dz^2} \mid = \frac{2 \zeta_{*}}{\Gamma x_{p}^2}
\approx \frac{2 \Phi_{*}}{\Gamma x_{p}^2}
\end{equation}
leads to the transformation from (\ref{h111}) - (\ref{h112})
to the following form

\begin{equation}\label{h115}
\Phi_{*} = \zeta_{*} - f_{*} c^{-4} \sum_{j=0}^{3} \frac{3!}{(3-j)!}
(-1)^{j} (-c x_{p})^{3-j} \exp(-cx_p) + f_{m} x_{p}^4 \lambda_{3}
\end{equation}

\begin{equation}\label{h116}
\frac{\Phi_{*}c}{\Gamma}
 = f_{*} c^{-3} 3 \sum_{j=0}^{2} \frac{2!}{(2-j)!}
(-1)^{j} (-c x_{p})^{2-j}
\exp(-c x_p)  + 3 f_{m} x_{p}^3 \lambda_{2}
\end{equation}

\begin{equation}\label{h117}
0 = -  f_{*} c^{-2} 6 \sum_{j=0}^{1}
(-1)^{j} (-c x_{p})^{1-j}
\exp(-c x_p)  + 6 f_{m} x_{p}^2 \lambda_{1} - \frac{2 \Phi_{*}}
{\Gamma x_{p}^2}
\end{equation}

From (\ref{h116}) it follows that
\begin{equation}\label{h118}
f_{*} =
\frac{\Phi_{*}c}{\Gamma}
[ c^{-3} 3 \sum_{j=0}^{2} \frac{2!}{(2-j)!}
(-1)^{j} (-c x_{p})^{2-j}
\exp(-c x_p)  + 3 \Psi x_{p}^3 \lambda_{2}]^{-1}
\end{equation}
where
$$
\Psi = \frac{f_m}{f_*}
$$
and (\ref{h117}) transforms to
\begin{equation}\label{h119}
- \frac{ \Phi_* c \{ c^{-2} 6 \sum_{j=0}^{1}
(-1)^{j} (-c x_{p})^{1-j} \exp(-c x_p)  + 6  x_{p}^2 \lambda_{1} \Psi \}
}
{
\Gamma
\{ c^{-3} 3 \sum_{j=0}^{2} \frac{2!}{(2-j)!}
(-1)^{j} (-c x_{p})^{2-j}\exp(-c x_p)  + 3 \Psi x_{p}^3 \lambda_{2} \}
}
 = \frac{2 \Phi_{*}}
{\Gamma x_{p}^2}
\end{equation}
To solve the last equation one can simplify it by the following transformation:
Consider $\Psi$ as some known value. It can be found  from (\ref{h115}).
In (\ref{h115}), (\ref{h116}) in some rough approximation the last
term is negligible
in comparison with the second term and one can spread the approximation of
the "ideal" supersaturation for all  this region. Then we immediately get
$$ \Psi = \exp(-1)$$
Another simplification is to notice that at the characteristic
scale $cx_p
\sim 1$
the value of  $\exp(-cx_p)$ is near to $\exp(-1)$ and can be linearized.
When $\Phi_{*}$ and $\Gamma$ are the given values then
equation (\ref{h119})  for
$x_{p}$ is the ordinary
algebraic equation of the power $4$
and can be analutically
solved. Then (\ref{h118}) gives the value for $f_{m}$ and, hence,
for $\Phi_{*}$ and $\Gamma$. This completes  the  current  step  of
the iteration
procedure for $\Phi_{*}$. The structure of solution in the nonlinear case is
the same one.

System (\ref{h115})-(\ref{h117}) allows some modifications which are
available for the nonlinear case also.
Note that the subintegral function $f(x)x$ in the expression for $d^2
g /dt^2$ is
essential only in the  finite region of the period of the
essential formation of the
droplets. The known approximation (\ref{h12}) also leads to the same conclusion.
Then  approximation (\ref{h12}) can be formally spread on the whole
region of the essential values of the function $f(x) x$. Then   equation
(\ref{h117}) leads to the following expression for $x_{p}$
\begin{equation}\label{hd17}
x_{p} = (\frac{2 \Phi_{*}}{3 \Gamma f_{m}} )^{1/4}
\end{equation}
In  the nonlinear case   equation (\ref{h117}) has the following form
\begin{equation}
\frac{d^2 \Phi }{dz^2} = 3 x_{p}^2 f_{m} - 2 \frac{\Phi_{*}}{\Gamma x_{p}^2}
\end{equation}
This bisquare equation gives $x_{p}$ dependence on $f_{m}$. Then the second
equation of  system (\ref{h115}) - (\ref{h117}) becomes the close
equation on the supersaturation and can be solved by iterations. These iterations
can utilize the sharp dependence of $\exp(-F_{c})$  on the supersaturation.
The zero approximation can be chosen as $\Gamma = 1, \Phi_{*} = 1$.
Then the second approximation gives rather a precise result for all parameters
of the spectrum. Certainly, the amplitude $f_{m}$ must be obtained by (\ref{h117})
instead of the  explicit expression from the
 classical theory of nucleation on the base of the supersaturation.

The system of the condensation equations can be even more simplified if one
notices that  approximation (\ref{h13}) can be spread
over the whole region of integration
in the first and in the
second equations of the system. This simplification is valid as far as
the subintegral functions in the first and in the second equations are
essential only in the "initial" region of
the period of the essential formation of droplets. Finally, the system
of the condensation equations has the following form

\begin{equation}\label{h221}
\frac{d\Phi}{dz}\mid_{z=0} =  3 f_{*} \int_{-\infty}^{0}
x^2 \exp(cx+lx^2) dx
\end{equation}

\begin{equation}\label{h222}
\frac{d^2 \Phi }{dz^2} = 6 \hat{\lambda}_{1}
 x_{p}^2 f_{m} - 2 \frac{\Phi_{*}}{\Gamma x_{p}^2}
\ \ \ \ \
\hat{\lambda}_{1} = \int_{-\infty}^{0} y \exp(-y^2) dy = 0.5
\end{equation}

One must take into account the difference between $f_{m}$ and $f_{*}$
which is given by

\begin{equation}\label{h223}
\frac{f_{m}}{f_{*}} = \exp(
-
\frac{\Gamma d\Phi}{3\Phi_* dz}\mid_{z=0}
\int_{-\infty}^{0}
x^3 \exp(cx+lx^2) dx / \int_{-\infty}^{0}
x^2 \exp(cx+lx^2) dx )
\end{equation}

It is essential that equations (\ref{h221}) and (\ref{h222}) are the
separate
ones. Equation\footnote{The integral can be calculated explicitly and
we have  the ordinary algebraic equation.}
(\ref{h221}) can be solved by iterations\footnote{As the algebraic equation
due to the sharp dependence of $f_*$ on the supersaturation}. The total number of
droplets can be obtained as

\begin{equation}
N = f_{m} \sqrt{\pi} x_{p} n_{\infty}
\end{equation}

The value of parameter $l$ is negative. But one can prove that
the relative error of the last expression $\delta$ increases when $l$
decreases in the absolute value
\begin{equation}
\frac{d \mid \delta \mid }{dl} > 0
\end{equation}

In the linear case

\begin{equation}
g(0) = \frac{\Phi_{*}}{\Gamma}
\end{equation}

\begin{equation}
x_{p} = c^{-1} (4\exp(1))^{1/4}
\end{equation}

\begin{equation}
N = \frac{\Phi_{*} c^3 (4\exp(1))^{1/4} \sqrt{\pi}}{6 \Gamma \exp(1)}
n_{\infty}
\end{equation}

Then $N$ is $1.19$ times greater than the result of iteration procedure
at the second step and $1.03$ times greater than the result
at the third step. It can be regarded as the practically precise result.
All mentioned above estimates obtained for $l=0$ remain valid in the
nonlinear case.

Figure 2 illustrates these constructions in the linear case.
This picture shows the contributions of the various forms of the spectrums
into the values of $g$, $dg/dz$, $d^2 g / d z^2$, $d^3 g / d z^3$
 (from upper till lower
axes of the coordinates) at\footnote{In the homogeneous condensation the
choice of $t_*$ corresponds to the maximum of the supersaturation.}
 $t=t_*$. Here the subingral expressions obtained in
the various approaches are drawn. The upper picture shows the subintegral
functions in the expression for $g$ at $z=0$, i.e. $x^3 f(x)$. The second
picture shows   the  subintegral function in the expression for $dg/3dz$
at $z=0$, i.e. $x^2 f(x)$. The third picture shows the behavior of the
subintegral expression for $d^2 g / 6 d z^2$ at $z=0$, i.e. $x f(x)$.
The lower picture shows the subintegral expression for the number of the
droplets or for  $d^3 g / 6 d z^3$, i.e. $f(x)$. For $f(x)$ several
different approaches are accepted. The curve "a" corresponds to the
precise universal solution, the curve "b" corresponds to the spectrum
calculated on the base of the ideal supersaturation (i.e. the
first iteration), the curve "c"
corresponds to the gaussian form of the spectrum and the curve "d" can
be drawn on the base of the spectrum of the gaussian type with the cube
correction term (see later).

It can be seen that for $g$ the suitable approximations can be given by
the  first iteration. For $dg/dz$ the suitable approximations can be given
by the first iteration and the  gaussian spectrum with a correction term
(with the appropriate cut-off). For $d^2g/dz^2$
and for $d^3 g/ dz^3$ the suitable results can be obtained from all
approximations except the first iteration.

The adopted  approximation has some resources of the
modification inside itself. Namely,
it can give  the value of the correction term and the value of the
boundary of the region of the applicability of this approximation.
 Really, let us add the next term
to the square approximation. Then we have
$$
f(x) \sim f_m \exp(-(\frac{x}{x_p})^2 -(\frac{x}{x_l})^3)
$$
with some parameter $x_l$. The value of $x_l$ will be established from the
fact that the model curve for the supersaturation
$$
\zeta_{appr} = \zeta_* - \frac{\zeta_*}{\Gamma}
(-(\frac{x}{x_p})^2 - (\frac{x}{x_l})^3)
$$
must only touch the line of the ideal supersaturation at some
point, which will mark the boundary between the initial region and the
region of the applicabily of above given approximation.

Figure 3 illustrates this construction.
Figure 3 demonstrates the form of the modified
spectrum with the cubic term taken into account.
The upper picture shows the behavior of the normalized deviation of the
supersaturation as function of $z$ for the different models.
The value $\delta_{id}$ corresponds to the ideal supersaturation, the
value $\delta_{sq}$ corresponds to the gaussian approximation, the
value $\delta_{mod}$ comes from the consideration of the gaussian with
a correction term.  The curve $\delta_{mod}$ touches the straight line
$\delta_{id}$. In the lower picture the gaussian spectrum $f_{gauss}$ and
the modified spectrum $f_{mod}$ are drawn. The scale of $x$ is chosen to
have $c=1$.  It is seen that the difference between these two spectrums is
rather small.

In the case of the linear source
it follows from the condition of equality of the derivatives of $\zeta_{appr}$
and $\Phi$ that
$$
x_l = (-\frac{2}{3} x_p^{-2}  x_b^{-1}  - \frac{1}{3} x_b^{-2} )^{-1/3}
$$
After the substitution of $x_l$
into the condition\footnote{We assume that approximately $\Psi = \exp(-1)$.}
$\zeta_{appr} (x_b) = \Phi(x_b) $
we shall come to
$$
x_b = \frac{ - 2 + ( 4 - 12 x_p^{-2})^{1/2} }
{2 x_p^{-2}}
$$
for the boundary between the initial and the central regions.

By the same way one can get the number of the molecules in the droplets by
the separate integration and, thus, fulfil the next step of the iteration
approximation.

\subsection{Heterogeneous condensation}

In the case of the heterogeneous condensation one can obtain the following
system at the period of the essential formation of the droplets

\begin{equation}\label{c1}
\Phi = \zeta + g
\ \ \ \ \ \ \
g(z) =
\int_{-\infty}^{z} dx (z-x)^3 f_{*}
\theta \exp(\frac{-\Gamma (\Phi_{*} -
\zeta)}{\Phi_{*}})
\end{equation}
\begin{equation}\label{c2}
\theta =
 \exp(-                    \frac{n_{\infty}}{\eta_{tot}}
\int_{-\infty}^{z} dx  f_{*}  \exp(\frac{-\Gamma (\Phi_{*} -
\zeta)}{\Phi_{*}})   )
\end{equation}

The straight development of the method utilized in the homogeneous case
leads to the construction of the square approximation for  the function

\begin{equation}\label{c3}
\psi = f_{*}
 \exp(-        \frac{n_{\infty}}{\eta_{tot}}
\int_{-\infty}^{z} dx  f_{*}  \exp(\frac{-\Gamma (\Phi_{*} -
\zeta(x) )}{\Phi_{*}}) )
 \exp(\frac{-\Gamma (\Phi_{*} -
\zeta(z))}{\Phi_{*}})
\end{equation}
which comes from the substitution of the second equation into the first one.

The square approximation for $\psi$ isn't suitable because the
factor $$
 \exp(-                 \frac{n_{\infty}}{\eta_{tot}}
\int_{-\infty}^{z} dx  f_{*}  \exp(\frac{-\Gamma (\Phi_{*} -
\zeta)}{\Phi_{*}})  )
$$
in the last expression has a very complex behavior. This term
goes from the expression for $\theta$. This allows to introduce the following
method. Notice that the iteration procedure in the investigation of
the heterogeneous condensation gives the appropriate result on the second step
\cite{Novosib}. The second approximation is obtained on the base of approximation
which doesn't contain any functional
dependence of $\theta$ upon $z$. Then one can solve the pseudo homogeneous
equation, i.e. equation (\ref{c1}) with $\theta \equiv 1$ and then calculate
$\theta$ according to the final step of the iteration procedure \cite{Novosib}.
The pseudo homogeneous equation will be solved by the method of the previous
subsections. This solution will give the main
 parameters $f_{m}, x_{p}$. The function
$\theta$ is approximately given by the following expression
\begin{equation}\label{c4}
\theta = \exp( - \Theta(z_{b}-z)\phi '  -
\Theta(z-z_{b}) \phi '')
\end{equation}
where
\begin{equation}
\phi ' = f_{*}  \frac{n_{\infty}}{\eta_{tot}}
\int_{-\infty}^{z} \exp(cx+lx^2) dx
\end{equation}
\begin{equation}
\phi '' =  f_{*}  \frac{n_{\infty}}{\eta_{tot}}
\int_{-\infty}^{z_{b}} \exp(cx+lx^2) dx +
f_{m}  \frac{n_{\infty}}{\eta_{tot}}
\int_{z_{b}}^{z} \exp(- (\frac{x}{x_{p}})^2) dx
\end{equation}
This expression can be taken in the  elementary functions due to
(\ref{Boyd}).

The number of the droplets which appears  from $\theta$
in a very simple form
\begin{equation}\label{c5}
N(z) = \eta_{tot}(1-\theta(z))
\end{equation}
  based on the conservation law for the heterogeneous centers.

One can introduce another approximation.
In the final formulas one can spread
 approximation
(\ref{h12}) over the whole
period of the essential formation of the droplets.
Then the value of $\theta_{final}$ can be presented in the  following
form
\begin{equation}
\theta_{final} = \exp[- \frac{n_{\infty}}{\eta_{tot}} f_{m} \sqrt{\pi} x_{p}]
\end{equation}
  and the total number of droplets is given by
\begin{equation}
N_{tot} = \eta_{tot}(1-\exp[- \frac{n_{\infty}}{\eta_{tot}}
 f_{m} \sqrt{\pi} x_{p}])
\end{equation}

The error of the above presented method is rather small,
the accuracy is practically the same as the
accuracy of the iteration procedure \cite{Novosib}. The advantage of this method lies
in the fact that the iterations can't be calculated in the case of the
nonlinear
 external conditions.

Note that in the above presented procedure the cross influence of
the exhaustion
of the heterogeneous centers can not be observed. Really, suppose that there
are  many  heterogeneous centers of the first sort and a few centers of
the other sort. Certainly,  the second sort can not act upon condensation
on the first sort in reality. But in final formulas the second sort acts
as there is no exhaustion of the second sort centers. The values $\Phi'$
and $\Phi''$ are based on some model for supersaturation. Only the parameter
$x_p$ is determined by the consumption of the vapor by the droplets. But it
is calculated on the base of the total values of the heterogeneous centers. So,
we must reconsider this procedure.

The process of exhaustion of
the heterogeneous centers makes the inequalities for the hierarchy
of the  regions even more strong.  So, we can repeat the above
presented procedure with the account of
the exhaustion of the heterogeneous centers.

 For the approximation of $\theta_j$ ($j$ marks the sort of the centers) in the
"initial" period one has
$$
\theta_j = \exp [ - \frac{f_{*\ j}}{c}
\frac{n_{\infty}}{\eta_{tot\ j}}
\exp(cx)
]
$$
Then this expression must be introduced into all expressions for $\alpha_i$.
These expressions now are the following ones
$$
\alpha_{i\ j} = (-1)^i \frac{3!}{i!(3-i)!} f_{* j}
\int_{-\infty}^{z_b} x^i \exp(cx+lx^2)
          \exp [ - \frac{f_{*\ j}}{c}  \frac{n_{\infty}}{\eta_{tot\ j}}
\exp(cx) ]
dx
$$

When the last expression is going to be calculated we shall also use the
steepens descent method and decompose $\theta$ into the Tailor's series near
$x=z_b$.

In the zero approximation
$$
\alpha_{i\ j} = (-1)^i \frac{3!}{i!(3-i)!} f_{* j}
\int_{-\infty}^{z_b} x^i \exp(cx+lx^2)
dx        \exp [ - \frac{f_{*\ j}}{c} \frac{n_{\infty}}{\eta_{tot\ j}}
\exp(cz_b) ]
$$
has the same analytical structure as the previous one. Also one can
decompose $\exp(cx)$ near $\exp(cz_b)$ and restrict this decomposition
by the zero and the
first two terms of the Tailor's serials.

The further generalization is rather simple. We must include $\sum_{j}$
all times when $f_{*\ j}$ appeared.

In the case of the continuous spectrum we must substitute $\sum_j$ by $\int
dw$ where $w$ is the activity of the heterogeneous center.

Note that the effect of  the cross influence of the exhaustion of
the heterogeneous centers
doesn't play any essential role in the dynamic conditions. This effect isn't
specific for dynamic conditions but takes place in the situation
of the decay type. The reason is that in the situation of the decay there are many
characteristic widths at one and the same (initial)
moment of time for the given initial value
of supersaturation. In dynamic conditions all halfwidths are approximately
equal to $c^{-1}$.

The alternative opportunity is to consider $\exp (l x^2)$
as the correction term and to keep
$\exp(-\frac{f_{* \ j}}{c} \exp(cx) )$
under the sign of integral.

One can say that in the situation of the dynamic conditions the process of
the exhaustion (the probability to be exhausted) occurs only due to the
activity
of this very sort of the nucleus (at the given supersaturation).

\section{Concluding remarks}

The applicability of the quasistationary distribution  as
the boundary
condition for the kinetic equation is based on the fact that the characteristic
time of the establishing of the stationary distribution in the region
$\nu^{1/3} \leq 3 \nu_c^{1/3}$ is negligible in comparison with the
characteristic
time of the variation of the supersaturation up to the value $\zeta /
\Gamma_i$. This statement is valid in all situations except the situations
when there occurs the total exhaustion of the free
heterogeneous centers and the
result is obvious - all centers of the given sort became the centers of the
droplets. This fact can be proved analytically. Also one can prove that
the main consumers of the vapor are the super-critical embryos, i.e. the
droplets.

The quasistationarity may not be valid when $\Delta x \gg x_p$. One can
show that in this case the effects of nonstationarity are cancelled.

Consider how to include  situation (\ref{noier}) in the general
consideration.
Note that in the iteration procedure \cite{Novosib} as far as in
the  modified method of the steepens descent
the structure of the calculations has one remarkable feature: the final
expressions
for $\theta$ and $N$ are given on the base of
the approximation obtained with
the help of the "ideal" laws. In these "ideal" laws the analytical structure
of the expressions for the different sorts is one and the same.
It turns to manifest
some features specific to the sort of the heterogeneous centers (due to their
number) only at the final iterations. So, the generalization is rather
formal and simple. It is necessary to have one common system of the coordinates
$x$, $z$ and  to calculate
directly the  iterations  (or to  use the  modified method
of the
steepens
descent). For $\theta_{(2)\ i}$ we get the separate
expressions:
$$
\theta_{(2)\ i} =
\exp(
- f_{*\ i} \frac{n_{\infty}}{\eta_{i\ tot}}
\frac{c^3 \Omega_{*}}{6 \Gamma_{i} (f_{*\ i} + f_{*\ j})}
(1 - \exp(
- \frac{6 \Gamma_{i}  (f_{*\ i} + f_{*\ j})}{c^4 \Omega_{*}}
\exp(cz)
)
)
)
$$
Here no cross influence of the exhaustion of the
heterogeneous centers of the different
sorts can be observed.
One must use the modified method of the steepens descent
 with the cross influence of
the exhaustion of the heterogeneous centers as had been made for the case
of the decay. The result of condensation on the spectrum of activities
will show that the form of the spectrum is near the universal one. So,
one can use it as some ansatz and formulate the algebraic equations on
the parameters.

The account of the nonisothermal effects is rather simple. The intensity of
the droplets formation is changed in $\frac{\zeta+1}{\zeta}
\ln(\zeta +1)$  times
(i.e. in the approximately constant number of
times during the period of the intensive formation
of the droplets). So, the description of the process can be attained
if we assume that the consumption of one molecule effectively leads to
the consumption of
$(k_2+1)$
molecules, where $k_2$
is some known parameter \cite{Novosib}.
So the description is reduced to the obvious renormalizations.

The global description of the evolution is more complex. One can
show that the already known procedure of the iteration approximations for the
external supersaturation  (where $z$ in all terms except $\frac{dz}{dt}$
is supposed to be known) is
successful\footnote{Here the initial approximation corresponds to
the ideal law of evolution} in this case also.
The application of this iteration procedure allows to  cancel the restriction
$k_2 \gg 1$
used in \cite{Novosib}.

Practically one needn't to fulfill such a complex procedure as to solve
many system of the integral equations many times. Ordinary the following
procedure must by accepted:
\begin{itemize}
\item
Put $\zeta = \Phi$ and determine the  moment $t_i$ of the exhaustion
of the different sorts of the heterogeneous centers.
\item
Establish the maximum of $\Omega$  under the monodisperse approximation
with the already defined coordinates of the peaks and $N_i = \eta_{tot\
i}$
\item
Reconsider formation of the droplets of that sorts of the heterogeneous
centers for which
$$
z(t_i) \geq z(t(max \  \Omega)) - \Delta x
$$
by the modified steepens descent method\footnote{Note that formation
of the droplets on the sorts for which
$$
z(t_i) \geq z(t(max \ \Omega )) + \Delta x
$$
can be treated in the pseudo homogeneous manner}.
\end{itemize}

As far as the moments of formation of the droplets of these sorts
are near $t(max\ \Omega)$ the action of formation can be presented
by the direct summation of all $f_{*\ i}$ for these sorts in the expression
for $g$ and its derivatives. The last step solves the problem.

\pagebreak
\begin{picture}(350,400)

\put(230,370){$t$}

\put(310,100){$f_1$}

\put(320,350){$f_2$}

\put(325,235){$f_3$}

\put(325,265){$f_4$}

\put(35,65){.}
\put(45,80){.}
\put(55,95){.}
\put(65,110){.}
\put(75,125){.}
\put(85,140){.}
\put(95,155){.}
\put(105,170){.}
\put(115,185){.}
\put(125,200){.}
\put(135,215){.}
\put(145,230){.}
\put(155,245){.}
\put(165,260){.}
\put(175,275){.}
\put(185,290){.}
\put(195,305){.}
\put(205,320){.}
\put(215,335){.}
\put(225,350){.}
\put(235,365){.}
\put(35,65){.}
\put(45,80){.}
\put(55,95){.}
\put(65,110){.}
\put(75,125){.}
\put(85,140){.}
\put(95,155){.}
\put(105,170){.}
\put(115,185){.}
\put(125,199){.}
\put(135,214){.}
\put(145,228){.}
\put(155,242){.}
\put(165,255){.}
\put(175,268){.}
\put(185,280){.}
\put(195,291){.}
\put(205,300){.}
\put(215,307){.}
\put(225,312){.}
\put(235,314){.}
\put(245,313){.}
\put(255,307){.}
\put(265,297){.}
\put(275,280){.}
\put(285,257){.}
\put(295,225){.}
\put(305,185){.}
\put(315,133){.}
\put(325,69){.}
\put(35,65){.}
\put(45,80){.}
\put(55,95){.}
\put(65,110){.}
\put(75,125){.}
\put(85,140){.}
\put(95,155){.}
\put(105,170){.}
\put(115,185){.}
\put(125,199){.}
\put(135,214){.}
\put(145,228){.}
\put(155,242){.}
\put(165,256){.}
\put(175,268){.}
\put(185,280){.}
\put(195,291){.}
\put(205,301){.}
\put(215,309){.}
\put(225,316){.}
\put(235,321){.}
\put(245,324){.}
\put(255,325){.}
\put(265,325){.}
\put(275,323){.}
\put(285,322){.}
\put(295,321){.}
\put(305,322){.}
\put(315,327){.}
\put(325,337){.}
\put(35,65){.}
\put(45,80){.}
\put(55,95){.}
\put(65,110){.}
\put(75,125){.}
\put(85,140){.}
\put(95,155){.}
\put(105,170){.}
\put(115,185){.}
\put(125,199){.}
\put(135,214){.}
\put(145,228){.}
\put(155,242){.}
\put(165,256){.}
\put(175,268){.}
\put(185,280){.}
\put(195,291){.}
\put(205,301){.}
\put(215,309){.}
\put(225,316){.}
\put(235,320){.}
\put(245,323){.}
\put(255,323){.}
\put(265,321){.}
\put(275,316){.}
\put(285,309){.}
\put(295,299){.}
\put(305,287){.}
\put(315,270){.}
\put(325,250){.}
\put(35,65){.}
\put(45,80){.}
\put(55,95){.}
\put(65,110){.}
\put(75,125){.}
\put(85,140){.}
\put(95,155){.}
\put(105,170){.}
\put(115,185){.}
\put(125,199){.}
\put(135,214){.}
\put(145,228){.}
\put(155,242){.}
\put(165,256){.}
\put(175,268){.}
\put(185,280){.}
\put(195,291){.}
\put(205,301){.}
\put(215,309){.}
\put(225,316){.}
\put(235,320){.}
\put(245,323){.}
\put(255,323){.}
\put(265,321){.}
\put(275,317){.}
\put(285,311){.}
\put(295,303){.}
\put(305,294){.}
\put(315,284){.}
\put(325,274){.}

\put(25,50){\vector(1,0){300}}
\put(25,40){0}
\put(225,40){1}
\put(320,40){$t$}
\put(150,20){$ Figure \  1 $}
\put(25,50){\vector(0,1){350}}
\put(30,350){1}
\put(30,380){$\zeta$}
\end{picture}

$$ Iterations \ for \   \Omega . $$
\pagebreak

\begin{picture}(350,470)
\put(2,50){.}
\put(2,150){.}
\put(2,251){.}
\put(2,359){.}
\put(9,50){.}
\put(9,150){.}
\put(9,251){.}
\put(9,360){.}
\put(17,50){.}
\put(17,150){.}
\put(17,251){.}
\put(17,360){.}
\put(24,50){.}
\put(24,150){.}
\put(24,252){.}
\put(24,361){.}
\put(32,50){.}
\put(32,150){.}
\put(32,252){.}
\put(32,363){.}
\put(39,50){.}
\put(39,150){.}
\put(39,252){.}
\put(39,364){.}
\put(47,50){.}
\put(47,150){.}
\put(47,252){.}
\put(47,365){.}
\put(54,50){.}
\put(54,150){.}
\put(54,252){.}
\put(54,366){.}
\put(62,50){.}
\put(62,150){.}
\put(62,253){.}
\put(62,368){.}
\put(69,50){.}
\put(69,150){.}
\put(69,253){.}
\put(69,369){.}
\put(77,50){.}
\put(77,150){.}
\put(77,253){.}
\put(77,371){.}
\put(84,50){.}
\put(84,151){.}
\put(84,254){.}
\put(84,373){.}
\put(92,50){.}
\put(92,151){.}
\put(92,254){.}
\put(92,374){.}
\put(99,50){.}
\put(99,151){.}
\put(99,254){.}
\put(99,376){.}
\put(107,50){.}
\put(107,151){.}
\put(107,255){.}
\put(107,378){.}
\put(114,50){.}
\put(114,151){.}
\put(114,255){.}
\put(114,380){.}
\put(122,50){.}
\put(122,151){.}
\put(122,256){.}
\put(122,383){.}
\put(129,50){.}
\put(129,151){.}
\put(129,256){.}
\put(129,385){.}
\put(137,50){.}
\put(137,151){.}
\put(137,257){.}
\put(137,387){.}
\put(144,50){.}
\put(144,152){.}
\put(144,258){.}
\put(144,390){.}
\put(152,50){.}
\put(152,152){.}
\put(152,258){.}
\put(152,392){.}
\put(159,50){.}
\put(159,152){.}
\put(159,259){.}
\put(159,395){.}
\put(167,50){.}
\put(167,152){.}
\put(167,260){.}
\put(167,397){.}
\put(174,51){.}
\put(174,152){.}
\put(174,261){.}
\put(174,399){.}
\put(182,51){.}
\put(182,153){.}
\put(182,262){.}
\put(182,402){.}
\put(189,51){.}
\put(189,153){.}
\put(189,263){.}
\put(189,404){.}
\put(197,51){.}
\put(197,153){.}
\put(197,264){.}
\put(197,406){.}
\put(204,51){.}
\put(204,154){.}
\put(204,265){.}
\put(204,408){.}
\put(212,51){.}
\put(212,154){.}
\put(212,266){.}
\put(212,410){.}
\put(219,51){.}
\put(219,155){.}
\put(219,267){.}
\put(219,411){.}
\put(227,52){.}
\put(227,155){.}
\put(227,268){.}
\put(227,412){.}
\put(234,52){.}
\put(234,156){.}
\put(234,269){.}
\put(234,413){.}
\put(242,52){.}
\put(242,156){.}
\put(242,270){.}
\put(242,413){.}
\put(249,52){.}
\put(249,157){.}
\put(249,271){.}
\put(249,413){.}
\put(257,53){.}
\put(257,158){.}
\put(257,272){.}
\put(257,412){.}
\put(264,53){.}
\put(264,158){.}
\put(264,272){.}
\put(264,411){.}
\put(272,54){.}
\put(272,159){.}
\put(272,273){.}
\put(272,409){.}
\put(279,54){.}
\put(279,160){.}
\put(279,273){.}
\put(279,406){.}
\put(287,55){.}
\put(287,160){.}
\put(287,274){.}
\put(287,403){.}
\put(294,55){.}
\put(294,161){.}
\put(294,273){.}
\put(294,399){.}
\put(302,56){.}
\put(302,162){.}
\put(302,273){.}
\put(302,395){.}
\put(309,57){.}
\put(309,162){.}
\put(309,272){.}
\put(309,390){.}
\put(317,58){.}
\put(317,163){.}
\put(317,271){.}
\put(317,385){.}
\put(324,59){.}
\put(324,163){.}
\put(324,270){.}
\put(324,380){.}
\put(332,60){.}
\put(332,163){.}
\put(332,268){.}
\put(332,374){.}
\put(339,61){.}
\put(339,163){.}
\put(339,266){.}
\put(339,369){.}
\put(347,62){.}
\put(347,162){.}
\put(347,263){.}
\put(347,364){.}
\put(354,63){.}
\put(354,162){.}
\put(354,261){.}
\put(354,360){.}
\put(362,64){.}
\put(362,161){.}
\put(362,258){.}
\put(362,356){.}
\put(369,65){.}
\put(369,159){.}
\put(369,256){.}
\put(369,353){.}
\put(377,66){.}
\put(377,157){.}
\put(377,253){.}
\put(377,352){.}
\put(384,67){.}
\put(384,155){.}
\put(384,252){.}
\put(384,350){.}
\put(392,67){.}
\put(392,153){.}
\put(392,250){.}
\put(392,350){.}
\put(399,67){.}
\put(399,150){.}
\put(399,250){.}
\put(399,350){.}
\put(2,50){.}
\put(2,150){.}
\put(2,251){.}
\put(2,359){.}
\put(9,50){.}
\put(9,150){.}
\put(9,251){.}
\put(9,360){.}
\put(17,50){.}
\put(17,150){.}
\put(17,251){.}
\put(17,361){.}
\put(24,50){.}
\put(24,150){.}
\put(24,252){.}
\put(24,362){.}
\put(32,50){.}
\put(32,150){.}
\put(32,252){.}
\put(32,363){.}
\put(39,50){.}
\put(39,150){.}
\put(39,252){.}
\put(39,364){.}
\put(47,50){.}
\put(47,150){.}
\put(47,252){.}
\put(47,365){.}
\put(54,50){.}
\put(54,150){.}
\put(54,252){.}
\put(54,366){.}
\put(62,50){.}
\put(62,150){.}
\put(62,253){.}
\put(62,368){.}
\put(69,50){.}
\put(69,150){.}
\put(69,253){.}
\put(69,369){.}
\put(77,50){.}
\put(77,151){.}
\put(77,253){.}
\put(77,371){.}
\put(84,50){.}
\put(84,151){.}
\put(84,254){.}
\put(84,373){.}
\put(92,50){.}
\put(92,151){.}
\put(92,254){.}
\put(92,375){.}
\put(99,50){.}
\put(99,151){.}
\put(99,254){.}
\put(99,377){.}
\put(107,50){.}
\put(107,151){.}
\put(107,255){.}
\put(107,379){.}
\put(114,50){.}
\put(114,151){.}
\put(114,255){.}
\put(114,381){.}
\put(122,50){.}
\put(122,151){.}
\put(122,256){.}
\put(122,383){.}
\put(129,50){.}
\put(129,151){.}
\put(129,257){.}
\put(129,385){.}
\put(137,50){.}
\put(137,151){.}
\put(137,257){.}
\put(137,388){.}
\put(144,50){.}
\put(144,152){.}
\put(144,258){.}
\put(144,390){.}
\put(152,50){.}
\put(152,152){.}
\put(152,259){.}
\put(152,393){.}
\put(159,50){.}
\put(159,152){.}
\put(159,259){.}
\put(159,395){.}
\put(167,50){.}
\put(167,152){.}
\put(167,260){.}
\put(167,398){.}
\put(174,51){.}
\put(174,152){.}
\put(174,261){.}
\put(174,400){.}
\put(182,51){.}
\put(182,153){.}
\put(182,262){.}
\put(182,403){.}
\put(189,51){.}
\put(189,153){.}
\put(189,263){.}
\put(189,405){.}
\put(197,51){.}
\put(197,154){.}
\put(197,264){.}
\put(197,408){.}
\put(204,51){.}
\put(204,154){.}
\put(204,265){.}
\put(204,410){.}
\put(212,51){.}
\put(212,154){.}
\put(212,266){.}
\put(212,412){.}
\put(219,51){.}
\put(219,155){.}
\put(219,268){.}
\put(219,414){.}
\put(227,52){.}
\put(227,155){.}
\put(227,269){.}
\put(227,415){.}
\put(234,52){.}
\put(234,156){.}
\put(234,270){.}
\put(234,416){.}
\put(242,52){.}
\put(242,157){.}
\put(242,271){.}
\put(242,417){.}
\put(249,52){.}
\put(249,157){.}
\put(249,272){.}
\put(249,417){.}
\put(257,53){.}
\put(257,158){.}
\put(257,273){.}
\put(257,417){.}
\put(264,53){.}
\put(264,159){.}
\put(264,274){.}
\put(264,416){.}
\put(272,54){.}
\put(272,160){.}
\put(272,275){.}
\put(272,415){.}
\put(279,54){.}
\put(279,161){.}
\put(279,276){.}
\put(279,413){.}
\put(287,55){.}
\put(287,162){.}
\put(287,277){.}
\put(287,410){.}
\put(294,56){.}
\put(294,163){.}
\put(294,277){.}
\put(294,407){.}
\put(302,57){.}
\put(302,164){.}
\put(302,277){.}
\put(302,403){.}
\put(309,58){.}
\put(309,165){.}
\put(309,277){.}
\put(309,399){.}
\put(317,60){.}
\put(317,166){.}
\put(317,276){.}
\put(317,393){.}
\put(324,61){.}
\put(324,167){.}
\put(324,275){.}
\put(324,388){.}
\put(332,63){.}
\put(332,167){.}
\put(332,274){.}
\put(332,382){.}
\put(339,65){.}
\put(339,168){.}
\put(339,272){.}
\put(339,376){.}
\put(347,67){.}
\put(347,168){.}
\put(347,269){.}
\put(347,371){.}
\put(354,70){.}
\put(354,168){.}
\put(354,267){.}
\put(354,365){.}
\put(362,73){.}
\put(362,168){.}
\put(362,264){.}
\put(362,360){.}
\put(369,77){.}
\put(369,167){.}
\put(369,260){.}
\put(369,356){.}
\put(377,82){.}
\put(377,165){.}
\put(377,257){.}
\put(377,353){.}
\put(384,87){.}
\put(384,161){.}
\put(384,254){.}
\put(384,351){.}
\put(392,93){.}
\put(392,157){.}
\put(392,251){.}
\put(392,350){.}
\put(399,100){.}
\put(399,150){.}
\put(399,250){.}
\put(399,350){.}
\put(2,50){.}
\put(2,150){.}
\put(2,250){.}
\put(2,350){.}
\put(9,50){.}
\put(9,150){.}
\put(9,250){.}
\put(9,350){.}
\put(17,50){.}
\put(17,150){.}
\put(17,250){.}
\put(17,350){.}
\put(24,50){.}
\put(24,150){.}
\put(24,250){.}
\put(24,350){.}
\put(32,50){.}
\put(32,150){.}
\put(32,250){.}
\put(32,350){.}
\put(39,50){.}
\put(39,150){.}
\put(39,250){.}
\put(39,350){.}
\put(47,50){.}
\put(47,150){.}
\put(47,250){.}
\put(47,350){.}
\put(54,50){.}
\put(54,150){.}
\put(54,250){.}
\put(54,350){.}
\put(62,50){.}
\put(62,150){.}
\put(62,250){.}
\put(62,350){.}
\put(69,50){.}
\put(69,150){.}
\put(69,250){.}
\put(69,350){.}
\put(77,50){.}
\put(77,150){.}
\put(77,250){.}
\put(77,350){.}
\put(84,50){.}
\put(84,150){.}
\put(84,250){.}
\put(84,350){.}
\put(92,50){.}
\put(92,150){.}
\put(92,250){.}
\put(92,350){.}
\put(99,50){.}
\put(99,150){.}
\put(99,250){.}
\put(99,350){.}
\put(107,50){.}
\put(107,150){.}
\put(107,250){.}
\put(107,350){.}
\put(114,50){.}
\put(114,150){.}
\put(114,250){.}
\put(114,350){.}
\put(122,50){.}
\put(122,150){.}
\put(122,250){.}
\put(122,350){.}
\put(129,50){.}
\put(129,150){.}
\put(129,250){.}
\put(129,350){.}
\put(137,50){.}
\put(137,150){.}
\put(137,250){.}
\put(137,351){.}
\put(144,50){.}
\put(144,150){.}
\put(144,250){.}
\put(144,351){.}
\put(152,50){.}
\put(152,150){.}
\put(152,250){.}
\put(152,351){.}
\put(159,50){.}
\put(159,150){.}
\put(159,250){.}
\put(159,352){.}
\put(167,50){.}
\put(167,150){.}
\put(167,251){.}
\put(167,353){.}
\put(174,50){.}
\put(174,150){.}
\put(174,251){.}
\put(174,354){.}
\put(182,50){.}
\put(182,150){.}
\put(182,251){.}
\put(182,355){.}
\put(189,50){.}
\put(189,150){.}
\put(189,252){.}
\put(189,356){.}
\put(197,50){.}
\put(197,151){.}
\put(197,252){.}
\put(197,358){.}
\put(204,50){.}
\put(204,151){.}
\put(204,253){.}
\put(204,361){.}
\put(212,50){.}
\put(212,151){.}
\put(212,254){.}
\put(212,363){.}
\put(219,50){.}
\put(219,151){.}
\put(219,255){.}
\put(219,367){.}
\put(227,50){.}
\put(227,152){.}
\put(227,256){.}
\put(227,370){.}
\put(234,51){.}
\put(234,152){.}
\put(234,257){.}
\put(234,374){.}
\put(242,51){.}
\put(242,153){.}
\put(242,259){.}
\put(242,378){.}
\put(249,51){.}
\put(249,154){.}
\put(249,261){.}
\put(249,382){.}
\put(257,52){.}
\put(257,154){.}
\put(257,263){.}
\put(257,386){.}
\put(264,52){.}
\put(264,155){.}
\put(264,265){.}
\put(264,389){.}
\put(272,53){.}
\put(272,156){.}
\put(272,267){.}
\put(272,392){.}
\put(279,53){.}
\put(279,158){.}
\put(279,268){.}
\put(279,394){.}
\put(287,54){.}
\put(287,159){.}
\put(287,270){.}
\put(287,395){.}
\put(294,55){.}
\put(294,160){.}
\put(294,271){.}
\put(294,395){.}
\put(302,56){.}
\put(302,161){.}
\put(302,272){.}
\put(302,393){.}
\put(309,57){.}
\put(309,162){.}
\put(309,272){.}
\put(309,390){.}
\put(317,58){.}
\put(317,163){.}
\put(317,272){.}
\put(317,386){.}
\put(324,59){.}
\put(324,164){.}
\put(324,271){.}
\put(324,382){.}
\put(332,60){.}
\put(332,164){.}
\put(332,269){.}
\put(332,376){.}
\put(339,62){.}
\put(339,164){.}
\put(339,267){.}
\put(339,371){.}
\put(347,63){.}
\put(347,164){.}
\put(347,265){.}
\put(347,366){.}
\put(354,64){.}
\put(354,163){.}
\put(354,262){.}
\put(354,361){.}
\put(362,65){.}
\put(362,162){.}
\put(362,259){.}
\put(362,357){.}
\put(369,66){.}
\put(369,160){.}
\put(369,256){.}
\put(369,354){.}
\put(377,67){.}
\put(377,158){.}
\put(377,254){.}
\put(377,352){.}
\put(384,68){.}
\put(384,156){.}
\put(384,252){.}
\put(384,351){.}
\put(392,68){.}
\put(392,153){.}
\put(392,250){.}
\put(392,350){.}
\put(399,68){.}
\put(399,150){.}
\put(399,250){.}
\put(399,350){.}
\put(2,50){.}
\put(2,150){.}
\put(2,252){.}
\put(2,363){.}
\put(9,50){.}
\put(9,150){.}
\put(9,252){.}
\put(9,362){.}
\put(17,50){.}
\put(17,150){.}
\put(17,252){.}
\put(17,362){.}
\put(24,50){.}
\put(24,150){.}
\put(24,252){.}
\put(24,362){.}
\put(32,50){.}
\put(32,150){.}
\put(32,252){.}
\put(32,362){.}
\put(39,50){.}
\put(39,150){.}
\put(39,252){.}
\put(39,362){.}
\put(47,50){.}
\put(47,150){.}
\put(47,252){.}
\put(47,362){.}
\put(54,50){.}
\put(54,150){.}
\put(54,252){.}
\put(54,362){.}
\put(62,50){.}
\put(62,150){.}
\put(62,252){.}
\put(62,362){.}
\put(69,50){.}
\put(69,150){.}
\put(69,252){.}
\put(69,363){.}
\put(77,50){.}
\put(77,150){.}
\put(77,252){.}
\put(77,363){.}
\put(84,50){.}
\put(84,150){.}
\put(84,252){.}
\put(84,364){.}
\put(92,50){.}
\put(92,150){.}
\put(92,252){.}
\put(92,365){.}
\put(99,50){.}
\put(99,150){.}
\put(99,253){.}
\put(99,366){.}
\put(107,50){.}
\put(107,150){.}
\put(107,253){.}
\put(107,367){.}
\put(114,50){.}
\put(114,151){.}
\put(114,253){.}
\put(114,368){.}
\put(122,50){.}
\put(122,151){.}
\put(122,254){.}
\put(122,370){.}
\put(129,50){.}
\put(129,151){.}
\put(129,254){.}
\put(129,371){.}
\put(137,50){.}
\put(137,151){.}
\put(137,254){.}
\put(137,373){.}
\put(144,50){.}
\put(144,151){.}
\put(144,255){.}
\put(144,375){.}
\put(152,50){.}
\put(152,151){.}
\put(152,256){.}
\put(152,377){.}
\put(159,50){.}
\put(159,151){.}
\put(159,256){.}
\put(159,380){.}
\put(167,50){.}
\put(167,151){.}
\put(167,257){.}
\put(167,382){.}
\put(174,50){.}
\put(174,152){.}
\put(174,258){.}
\put(174,385){.}
\put(182,50){.}
\put(182,152){.}
\put(182,259){.}
\put(182,388){.}
\put(189,51){.}
\put(189,152){.}
\put(189,260){.}
\put(189,391){.}
\put(197,51){.}
\put(197,153){.}
\put(197,261){.}
\put(197,394){.}
\put(204,51){.}
\put(204,153){.}
\put(204,262){.}
\put(204,398){.}
\put(212,51){.}
\put(212,154){.}
\put(212,264){.}
\put(212,401){.}
\put(219,51){.}
\put(219,154){.}
\put(219,265){.}
\put(219,404){.}
\put(227,51){.}
\put(227,155){.}
\put(227,266){.}
\put(227,407){.}
\put(234,52){.}
\put(234,155){.}
\put(234,268){.}
\put(234,410){.}
\put(242,52){.}
\put(242,156){.}
\put(242,270){.}
\put(242,412){.}
\put(249,52){.}
\put(249,157){.}
\put(249,271){.}
\put(249,414){.}
\put(257,53){.}
\put(257,158){.}
\put(257,273){.}
\put(257,415){.}
\put(264,53){.}
\put(264,159){.}
\put(264,274){.}
\put(264,415){.}
\put(272,54){.}
\put(272,160){.}
\put(272,275){.}
\put(272,414){.}
\put(279,54){.}
\put(279,161){.}
\put(279,276){.}
\put(279,413){.}
\put(287,55){.}
\put(287,162){.}
\put(287,277){.}
\put(287,410){.}
\put(294,56){.}
\put(294,163){.}
\put(294,277){.}
\put(294,407){.}
\put(302,57){.}
\put(302,164){.}
\put(302,277){.}
\put(302,402){.}
\put(309,58){.}
\put(309,164){.}
\put(309,276){.}
\put(309,397){.}
\put(317,59){.}
\put(317,165){.}
\put(317,275){.}
\put(317,391){.}
\put(324,60){.}
\put(324,165){.}
\put(324,273){.}
\put(324,385){.}
\put(332,61){.}
\put(332,165){.}
\put(332,271){.}
\put(332,378){.}
\put(339,62){.}
\put(339,165){.}
\put(339,268){.}
\put(339,372){.}
\put(347,63){.}
\put(347,164){.}
\put(347,265){.}
\put(347,366){.}
\put(354,65){.}
\put(354,163){.}
\put(354,262){.}
\put(354,361){.}
\put(362,66){.}
\put(362,162){.}
\put(362,259){.}
\put(362,357){.}
\put(369,67){.}
\put(369,160){.}
\put(369,256){.}
\put(369,354){.}
\put(377,67){.}
\put(377,158){.}
\put(377,254){.}
\put(377,352){.}
\put(384,68){.}
\put(384,156){.}
\put(384,252){.}
\put(384,351){.}
\put(392,68){.}
\put(392,153){.}
\put(392,250){.}
\put(392,350){.}
\put(399,68){.}
\put(399,150){.}
\put(399,250){.}
\put(399,350){.}
\put(390,70){a=c=d}
\put(390,105){b}
\put(340,152){a=c=d}
\put(275,152){a=b=c}
\put(235,160){d}
\put(380,165){b}
\put(2,240){a=b=c=d}
\put(265,257){c}
\put(300,265){a=c}
\put(300,285){b=d}
\put(335,252){a=c=d}
\put(350,275){b}
\put(2,370){d}
\put(150,405){a=b}
\put(150,370){d}
\put(265,380){c}
\put(265,399){a}
\put(10,50){\vector(1,0){400}}
\put(10,150){\vector(1,0){400}}
\put(10,250){\vector(1,0){400}}
\put(10,350){\vector(1,0){400}}
\put(400,35){0}
\put(350,35){-1}
\put(410,35){$x$}
\put(400,135){0}
\put(350,135){-1}
\put(410,135){$x$}
\put(400,235){0}
\put(350,235){-1}
\put(410,235){$x$}
\put(400,335){0}
\put(350,335){-1}
\put(410,335){$x$}
\put(150,10){$Figure \  2$ }

\end{picture}
$$ Subintegral \ functions \ in \ different \ approximations. $$

\pagebreak

\begin{picture}(320,500)
\put(95,51){.}
\put(95,52){.}
\put(95,53){.}
\put(95,35){$x_b$}
\put(13,51){.}
\put(13,56){.}
\put(13,242){.}
\put(13,253){.}
\put(17,52){.}
\put(17,57){.}
\put(17,247){.}
\put(17,257){.}
\put(21,52){.}
\put(21,58){.}
\put(21,251){.}
\put(21,261){.}
\put(25,52){.}
\put(25,58){.}
\put(25,256){.}
\put(25,265){.}
\put(29,53){.}
\put(29,59){.}
\put(29,202){.}
\put(29,261){.}
\put(29,269){.}
\put(33,53){.}
\put(33,60){.}
\put(33,211){.}
\put(33,266){.}
\put(33,273){.}
\put(37,54){.}
\put(37,61){.}
\put(37,220){.}
\put(37,271){.}
\put(37,277){.}
\put(41,55){.}
\put(41,62){.}
\put(41,228){.}
\put(41,276){.}
\put(41,281){.}
\put(45,56){.}
\put(45,64){.}
\put(45,236){.}
\put(45,280){.}
\put(45,285){.}
\put(49,57){.}
\put(49,65){.}
\put(49,244){.}
\put(49,285){.}
\put(49,289){.}
\put(53,58){.}
\put(53,67){.}
\put(53,251){.}
\put(53,290){.}
\put(53,293){.}
\put(57,59){.}
\put(57,68){.}
\put(57,259){.}
\put(57,294){.}
\put(57,297){.}
\put(61,60){.}
\put(61,70){.}
\put(61,266){.}
\put(61,299){.}
\put(61,301){.}
\put(65,62){.}
\put(65,72){.}
\put(65,273){.}
\put(65,304){.}
\put(65,305){.}
\put(69,64){.}
\put(69,74){.}
\put(69,280){.}
\put(69,308){.}
\put(69,309){.}
\put(73,66){.}
\put(73,76){.}
\put(73,287){.}
\put(73,312){.}
\put(73,313){.}
\put(77,68){.}
\put(77,78){.}
\put(77,294){.}
\put(77,317){.}
\put(77,317){.}
\put(81,70){.}
\put(81,81){.}
\put(81,300){.}
\put(81,321){.}
\put(81,321){.}
\put(85,73){.}
\put(85,84){.}
\put(85,306){.}
\put(85,325){.}
\put(85,325){.}
\put(89,76){.}
\put(89,87){.}
\put(89,312){.}
\put(89,329){.}
\put(89,329){.}
\put(93,79){.}
\put(93,90){.}
\put(93,318){.}
\put(93,333){.}
\put(93,333){.}
\put(97,82){.}
\put(97,93){.}
\put(97,323){.}
\put(97,337){.}
\put(97,337){.}
\put(101,86){.}
\put(101,96){.}
\put(101,329){.}
\put(101,341){.}
\put(101,341){.}
\put(105,90){.}
\put(105,100){.}
\put(105,334){.}
\put(105,345){.}
\put(105,345){.}
\put(109,94){.}
\put(109,104){.}
\put(109,339){.}
\put(109,349){.}
\put(109,349){.}
\put(113,98){.}
\put(113,108){.}
\put(113,344){.}
\put(113,353){.}
\put(113,353){.}
\put(117,103){.}
\put(117,112){.}
\put(117,348){.}
\put(117,356){.}
\put(117,357){.}
\put(121,108){.}
\put(121,117){.}
\put(121,352){.}
\put(121,359){.}
\put(121,361){.}
\put(125,113){.}
\put(125,121){.}
\put(125,357){.}
\put(125,363){.}
\put(125,365){.}
\put(129,118){.}
\put(129,126){.}
\put(129,361){.}
\put(129,366){.}
\put(129,369){.}
\put(133,124){.}
\put(133,131){.}
\put(133,365){.}
\put(133,369){.}
\put(133,373){.}
\put(137,129){.}
\put(137,136){.}
\put(137,368){.}
\put(137,372){.}
\put(137,377){.}
\put(141,135){.}
\put(141,141){.}
\put(141,372){.}
\put(141,375){.}
\put(141,381){.}
\put(145,141){.}
\put(145,146){.}
\put(145,375){.}
\put(145,377){.}
\put(145,385){.}
\put(149,146){.}
\put(149,151){.}
\put(149,378){.}
\put(149,380){.}
\put(149,389){.}
\put(153,152){.}
\put(153,156){.}
\put(153,381){.}
\put(153,382){.}
\put(153,393){.}
\put(157,157){.}
\put(157,161){.}
\put(157,383){.}
\put(157,385){.}
\put(157,397){.}
\put(161,163){.}
\put(161,165){.}
\put(161,386){.}
\put(161,387){.}
\put(161,401){.}
\put(165,168){.}
\put(165,170){.}
\put(165,388){.}
\put(165,389){.}
\put(165,405){.}
\put(169,173){.}
\put(169,175){.}
\put(169,390){.}
\put(169,391){.}
\put(169,409){.}
\put(173,178){.}
\put(173,179){.}
\put(173,392){.}
\put(173,392){.}
\put(173,413){.}
\put(177,182){.}
\put(177,183){.}
\put(177,394){.}
\put(177,394){.}
\put(177,417){.}
\put(181,186){.}
\put(181,187){.}
\put(181,395){.}
\put(181,395){.}
\put(181,421){.}
\put(185,189){.}
\put(185,190){.}
\put(185,396){.}
\put(185,397){.}
\put(185,425){.}
\put(189,193){.}
\put(189,193){.}
\put(189,397){.}
\put(189,398){.}
\put(189,429){.}
\put(193,195){.}
\put(193,195){.}
\put(193,398){.}
\put(193,398){.}
\put(193,433){.}
\put(197,197){.}
\put(197,197){.}
\put(197,399){.}
\put(197,399){.}
\put(197,437){.}
\put(201,199){.}
\put(201,199){.}
\put(201,400){.}
\put(201,400){.}
\put(201,441){.}
\put(205,200){.}
\put(205,200){.}
\put(205,400){.}
\put(205,400){.}
\put(205,445){.}
\put(209,200){.}
\put(209,200){.}
\put(209,400){.}
\put(209,400){.}
\put(209,449){.}
\put(213,200){.}
\put(213,200){.}
\put(213,400){.}
\put(213,400){.}
\put(213,453){.}
\put(217,199){.}
\put(217,199){.}
\put(217,400){.}
\put(217,400){.}
\put(217,457){.}
\put(221,198){.}
\put(221,198){.}
\put(221,399){.}
\put(221,399){.}
\put(221,461){.}
\put(225,196){.}
\put(225,196){.}
\put(225,399){.}
\put(225,399){.}
\put(225,465){.}
\put(229,193){.}
\put(229,193){.}
\put(229,398){.}
\put(229,398){.}
\put(229,469){.}
\put(233,190){.}
\put(233,190){.}
\put(233,397){.}
\put(233,397){.}
\put(233,473){.}
\put(237,187){.}
\put(237,186){.}
\put(237,395){.}
\put(237,395){.}
\put(241,183){.}
\put(241,182){.}
\put(241,394){.}
\put(241,394){.}
\put(245,179){.}
\put(245,178){.}
\put(245,392){.}
\put(245,392){.}
\put(249,174){.}
\put(249,173){.}
\put(249,391){.}
\put(249,390){.}
\put(253,169){.}
\put(253,167){.}
\put(253,389){.}
\put(253,388){.}
\put(257,164){.}
\put(257,162){.}
\put(257,386){.}
\put(257,385){.}
\put(261,159){.}
\put(261,156){.}
\put(261,384){.}
\put(261,383){.}
\put(265,153){.}
\put(265,150){.}
\put(265,381){.}
\put(265,380){.}
\put(269,148){.}
\put(269,144){.}
\put(269,379){.}
\put(269,376){.}
\put(273,142){.}
\put(273,137){.}
\put(273,376){.}
\put(273,373){.}
\put(277,136){.}
\put(277,131){.}
\put(277,372){.}
\put(277,369){.}
\put(281,131){.}
\put(281,125){.}
\put(281,369){.}
\put(281,365){.}
\put(285,125){.}
\put(285,119){.}
\put(285,365){.}
\put(285,361){.}
\put(289,120){.}
\put(289,113){.}
\put(289,362){.}
\put(289,357){.}
\put(293,114){.}
\put(293,107){.}
\put(293,358){.}
\put(293,352){.}
\put(297,109){.}
\put(297,102){.}
\put(297,354){.}
\put(297,347){.}
\put(301,104){.}
\put(301,97){.}
\put(301,349){.}
\put(301,342){.}
\put(305,100){.}
\put(305,92){.}
\put(305,345){.}
\put(305,336){.}
\put(309,95){.}
\put(309,87){.}
\put(309,340){.}
\put(309,330){.}
\put(313,91){.}
\put(313,83){.}
\put(313,335){.}
\put(313,324){.}
\put(317,87){.}
\put(317,79){.}
\put(317,330){.}
\put(317,317){.}
\put(321,83){.}
\put(321,75){.}
\put(321,325){.}
\put(321,311){.}
\put(325,80){.}
\put(325,72){.}
\put(325,319){.}
\put(325,304){.}
\put(329,77){.}
\put(329,69){.}
\put(329,313){.}
\put(329,296){.}
\put(333,74){.}
\put(333,66){.}
\put(333,307){.}
\put(333,289){.}
\put(337,71){.}
\put(337,64){.}
\put(337,301){.}
\put(337,281){.}
\put(341,68){.}
\put(341,62){.}
\put(341,295){.}
\put(341,272){.}
\put(345,66){.}
\put(345,60){.}
\put(345,289){.}
\put(345,264){.}
\put(349,64){.}
\put(349,58){.}
\put(349,282){.}
\put(349,255){.}
\put(353,62){.}
\put(353,57){.}
\put(353,275){.}
\put(353,246){.}
\put(357,61){.}
\put(357,56){.}
\put(357,268){.}
\put(357,236){.}
\put(361,59){.}
\put(361,55){.}
\put(361,261){.}
\put(361,226){.}
\put(365,58){.}
\put(365,54){.}
\put(365,253){.}
\put(365,216){.}
\put(369,57){.}
\put(369,53){.}
\put(369,246){.}
\put(369,205){.}
\put(373,56){.}
\put(373,52){.}
\put(373,238){.}
\put(377,55){.}
\put(377,52){.}
\put(377,230){.}
\put(381,54){.}
\put(381,52){.}
\put(381,222){.}
\put(385,54){.}
\put(385,51){.}
\put(385,213){.}
\put(389,53){.}
\put(389,51){.}
\put(389,205){.}
\put(393,53){.}
\put(393,51){.}
\put(397,52){.}
\put(397,51){.}
\put(401,52){.}
\put(401,50){.}
\put(405,51){.}
\put(405,50){.}
\put(409,51){.}
\put(409,50){.}
\put(10,50){\vector(1,0){400}}
\put(210,35){0}
\put(260,35){1}
\put(390,35){$z$}
\put(10,450){\vector(1,0){400}}
\put(210,240){\vector(0,1){250}}
\put(220,485){$\delta = \frac{\Gamma (\zeta(z) - \Phi_*) }{ \Phi_*} $}
\put(230,460){$\delta_{id}$}
\put(22,246){$\delta_{mod}$}
\put(3,260){$\delta_{id}$}
\put(300,300){$\delta_{mod}$}
\put(70,270){$\delta_{sq}$}
\put(350,300){$\delta_{sq}$}
\put(220,250){-4}
\put(220,300){-3}
\put(220,350){-2}
\put(220,400){-1}

\put(30,85){$f_{mod}$}
\put(120,85){$f_{gauss}$}
\put(270,85){$f_{mod}$}
\put(350,85){$f_{gauss}$}

\put(220,435){0}
\put(270,435){1}
\put(390,435){$z$}
\put(170,10){$Figure \  3$ }

\end{picture}

$$ Modified \ spectrum \ with \ cubic \ term. $$

\end{document}